\documentclass[longauth]{aaEC}

\usepackage{graphicx}
\usepackage{natbib}
\usepackage{scalerel}
\usepackage{siunitx}
\usepackage{lastpage}
\usepackage{soul}
\usepackage{amsmath}

\usepackage[table]{xcolor}

\newcolumntype{P}[1]{>{\centering\arraybackslash}p{#1}}
\newcommand*{\rom}[1]{\expandafter\@slowromancap\romannumeral #1@}
\newcommand{\orcid}[1]{} 

\usepackage{txfonts}
\usepackage[pdfencoding=auto,psdextra]{hyperref}
\hypersetup{
    colorlinks=true,
    linkcolor=blue,
    filecolor=magenta,      
    urlcolor=blue,
    citecolor=blue
}
\newcommand{\rev}{}

\makeatletter
\renewcommand*\aa@pageof{, page \thepage{} of \pageref*{LastPage}}
\makeatother

\usepackage[utf8]{inputenc}
\usepackage{datetime2}

\usepackage[switch, modulo]{lineno}

\usepackage{euclid}

\begin{document}

%
%

\title{Euclid Quick Data Release (Q1)}
\subtitle{Quenching precedes bulge formation in dense environments but follows it in the field}

\renewcommand{\orcid}[1]{} 
\author{Euclid Collaboration: F.~Gentile\orcid{0000-0002-8008-9871}\thanks{\email{fabrizio.gentile@cea.fr}}\inst{\ref{aff1},\ref{aff2}}
\and E.~Daddi\orcid{0000-0002-3331-9590}\inst{\ref{aff3}}
\and D.~Elbaz\orcid{0000-0002-7631-647X}\inst{\ref{aff3}}
\and A.~Enia\orcid{0000-0002-0200-2857}\inst{\ref{aff4},\ref{aff2}}
\and B.~Magnelli\orcid{0000-0002-6777-6490}\inst{\ref{aff3}}
\and J-B.~Billand\orcid{0009-0004-4168-3634}\inst{\ref{aff1}}
\and P.~Corcho-Caballero\orcid{0000-0001-6327-7080}\inst{\ref{aff5}}
\and C.~Cleland\orcid{0009-0002-1769-1437}\inst{\ref{aff6}}
\and G.~De~Lucia\orcid{0000-0002-6220-9104}\inst{\ref{aff7}}
\and C.~D'Eugenio\orcid{0000-0001-7344-3126}\inst{\ref{aff8},\ref{aff1}}
\and M.~Fossati\orcid{0000-0002-9043-8764}\inst{\ref{aff9},\ref{aff10}}
\and M.~Franco\orcid{0000-0002-3560-8599}\inst{\ref{aff3}}
\and C.~Lobo\orcid{0000-0003-2415-3338}\inst{\ref{aff11},\ref{aff12}}
\and Y.~Lyu\orcid{0000-0003-0355-0633}\inst{\ref{aff1}}
\and M.~Magliocchetti\orcid{0000-0001-9158-4838}\inst{\ref{aff13}}
\and G.~A.~Mamon\orcid{0000-0001-8956-5953}\inst{\ref{aff14},\ref{aff15}}
\and L.~Quilley\orcid{0009-0008-8375-8605}\inst{\ref{aff16}}
\and J.~G.~Sorce\orcid{0000-0002-2307-2432}\inst{\ref{aff17},\ref{aff18}}
\and M.~Tarrasse\orcid{0009-0009-3123-4479}\inst{\ref{aff1}}
\and M.~Bolzonella\orcid{0000-0003-3278-4607}\inst{\ref{aff2}}
\and F.~Durret\orcid{0000-0002-6991-4578}\inst{\ref{aff15}}
\and L.~Gabarra\orcid{0000-0002-8486-8856}\inst{\ref{aff19}}
\and S.~Guo\inst{\ref{aff1}}
\and L.~Pozzetti\orcid{0000-0001-7085-0412}\inst{\ref{aff2}}
\and S.~Quai\orcid{0000-0002-0449-8163}\inst{\ref{aff20},\ref{aff2}}
\and F.~Shankar\orcid{0000-0001-8973-5051}\inst{\ref{aff21}}
\and V.~Sangalli\inst{\ref{aff1},\ref{aff3}}
\and M.~Talia\orcid{0000-0003-4352-2063}\inst{\ref{aff20},\ref{aff2}}
\and M.~Baes\orcid{0000-0002-3930-2757}\inst{\ref{aff22}}
\and H.~Fu\orcid{0009-0002-8051-1056}\inst{\ref{aff23},\ref{aff21}}
\and M.~Girardi\orcid{0000-0003-1861-1865}\inst{\ref{aff24},\ref{aff7}}
\and J.~Matthee\orcid{0000-0003-2871-127X}\inst{\ref{aff25}}
\and P.~A.~Oesch\orcid{0000-0001-5851-6649}\inst{\ref{aff26},\ref{aff27},\ref{aff28}}
\and D.~Roberts\orcid{0009-0009-7662-0445}\inst{\ref{aff21}}
\and J.~Schaye\orcid{0000-0002-0668-5560}\inst{\ref{aff29}}
\and D.~Scott\orcid{0000-0002-6878-9840}\inst{\ref{aff30}}
\and L.~Spinoglio\orcid{0000-0001-8840-1551}\inst{\ref{aff13}}
\and B.~Altieri\orcid{0000-0003-3936-0284}\inst{\ref{aff31}}
\and A.~Amara\inst{\ref{aff32}}
\and S.~Andreon\orcid{0000-0002-2041-8784}\inst{\ref{aff10}}
\and N.~Auricchio\orcid{0000-0003-4444-8651}\inst{\ref{aff2}}
\and C.~Baccigalupi\orcid{0000-0002-8211-1630}\inst{\ref{aff33},\ref{aff7},\ref{aff34},\ref{aff35}}
\and M.~Baldi\orcid{0000-0003-4145-1943}\inst{\ref{aff4},\ref{aff2},\ref{aff36}}
\and A.~Balestra\orcid{0000-0002-6967-261X}\inst{\ref{aff37}}
\and S.~Bardelli\orcid{0000-0002-8900-0298}\inst{\ref{aff2}}
\and R.~Bender\orcid{0000-0001-7179-0626}\inst{\ref{aff38},\ref{aff39}}
\and A.~Biviano\orcid{0000-0002-0857-0732}\inst{\ref{aff7},\ref{aff33}}
\and E.~Branchini\orcid{0000-0002-0808-6908}\inst{\ref{aff40},\ref{aff41},\ref{aff10}}
\and M.~Brescia\orcid{0000-0001-9506-5680}\inst{\ref{aff42},\ref{aff43}}
\and J.~Brinchmann\orcid{0000-0003-4359-8797}\inst{\ref{aff11},\ref{aff44},\ref{aff45}}
\and S.~Camera\orcid{0000-0003-3399-3574}\inst{\ref{aff46},\ref{aff47},\ref{aff48}}
\and G.~Ca\~nas-Herrera\orcid{0000-0003-2796-2149}\inst{\ref{aff49},\ref{aff29}}
\and V.~Capobianco\orcid{0000-0002-3309-7692}\inst{\ref{aff48}}
\and C.~Carbone\orcid{0000-0003-0125-3563}\inst{\ref{aff50}}
\and J.~Carretero\orcid{0000-0002-3130-0204}\inst{\ref{aff51},\ref{aff52}}
\and S.~Casas\orcid{0000-0002-4751-5138}\inst{\ref{aff53},\ref{aff54}}
\and M.~Castellano\orcid{0000-0001-9875-8263}\inst{\ref{aff55}}
\and G.~Castignani\orcid{0000-0001-6831-0687}\inst{\ref{aff2}}
\and S.~Cavuoti\orcid{0000-0002-3787-4196}\inst{\ref{aff43},\ref{aff56}}
\and K.~C.~Chambers\orcid{0000-0001-6965-7789}\inst{\ref{aff57}}
\and A.~Cimatti\inst{\ref{aff58}}
\and C.~Colodro-Conde\inst{\ref{aff59}}
\and G.~Congedo\orcid{0000-0003-2508-0046}\inst{\ref{aff60}}
\and L.~Conversi\orcid{0000-0002-6710-8476}\inst{\ref{aff61},\ref{aff31}}
\and Y.~Copin\orcid{0000-0002-5317-7518}\inst{\ref{aff62}}
\and F.~Courbin\orcid{0000-0003-0758-6510}\inst{\ref{aff63},\ref{aff64}}
\and H.~M.~Courtois\orcid{0000-0003-0509-1776}\inst{\ref{aff65}}
\and M.~Cropper\orcid{0000-0003-4571-9468}\inst{\ref{aff66}}
\and A.~Da~Silva\orcid{0000-0002-6385-1609}\inst{\ref{aff67},\ref{aff68}}
\and H.~Degaudenzi\orcid{0000-0002-5887-6799}\inst{\ref{aff26}}
\and C.~Dolding\orcid{0009-0003-7199-6108}\inst{\ref{aff66}}
\and H.~Dole\orcid{0000-0002-9767-3839}\inst{\ref{aff18}}
\and F.~Dubath\orcid{0000-0002-6533-2810}\inst{\ref{aff26}}
\and C.~A.~J.~Duncan\orcid{0009-0003-3573-0791}\inst{\ref{aff60}}
\and X.~Dupac\inst{\ref{aff31}}
\and S.~Dusini\orcid{0000-0002-1128-0664}\inst{\ref{aff69}}
\and S.~Escoffier\orcid{0000-0002-2847-7498}\inst{\ref{aff70}}
\and M.~Fabricius\orcid{0000-0002-7025-6058}\inst{\ref{aff38},\ref{aff39}}
\and M.~Farina\orcid{0000-0002-3089-7846}\inst{\ref{aff13}}
\and R.~Farinelli\inst{\ref{aff2}}
\and S.~Ferriol\inst{\ref{aff62}}
\and F.~Finelli\orcid{0000-0002-6694-3269}\inst{\ref{aff2},\ref{aff71}}
\and N.~Fourmanoit\orcid{0009-0005-6816-6925}\inst{\ref{aff70}}
\and M.~Frailis\orcid{0000-0002-7400-2135}\inst{\ref{aff7}}
\and E.~Franceschi\orcid{0000-0002-0585-6591}\inst{\ref{aff2}}
\and M.~Fumana\orcid{0000-0001-6787-5950}\inst{\ref{aff50}}
\and S.~Galeotta\orcid{0000-0002-3748-5115}\inst{\ref{aff7}}
\and K.~George\orcid{0000-0002-1734-8455}\inst{\ref{aff72}}
\and B.~Gillis\orcid{0000-0002-4478-1270}\inst{\ref{aff60}}
\and C.~Giocoli\orcid{0000-0002-9590-7961}\inst{\ref{aff2},\ref{aff36}}
\and J.~Gracia-Carpio\inst{\ref{aff38}}
\and A.~Grazian\orcid{0000-0002-5688-0663}\inst{\ref{aff37}}
\and F.~Grupp\inst{\ref{aff38},\ref{aff39}}
\and S.~Gwyn\orcid{0000-0001-8221-8406}\inst{\ref{aff73}}
\and S.~V.~H.~Haugan\orcid{0000-0001-9648-7260}\inst{\ref{aff74}}
\and J.~Hoar\inst{\ref{aff31}}
\and W.~Holmes\inst{\ref{aff75}}
\and I.~M.~Hook\orcid{0000-0002-2960-978X}\inst{\ref{aff76}}
\and F.~Hormuth\inst{\ref{aff77}}
\and A.~Hornstrup\orcid{0000-0002-3363-0936}\inst{\ref{aff78},\ref{aff79}}
\and K.~Jahnke\orcid{0000-0003-3804-2137}\inst{\ref{aff80}}
\and M.~Jhabvala\inst{\ref{aff81}}
\and B.~Joachimi\orcid{0000-0001-7494-1303}\inst{\ref{aff82}}
\and E.~Keih\"anen\orcid{0000-0003-1804-7715}\inst{\ref{aff83}}
\and S.~Kermiche\orcid{0000-0002-0302-5735}\inst{\ref{aff70}}
\and A.~Kiessling\orcid{0000-0002-2590-1273}\inst{\ref{aff75}}
\and B.~Kubik\orcid{0009-0006-5823-4880}\inst{\ref{aff62}}
\and M.~K\"ummel\orcid{0000-0003-2791-2117}\inst{\ref{aff39}}
\and M.~Kunz\orcid{0000-0002-3052-7394}\inst{\ref{aff84}}
\and H.~Kurki-Suonio\orcid{0000-0002-4618-3063}\inst{\ref{aff85},\ref{aff86}}
\and A.~M.~C.~Le~Brun\orcid{0000-0002-0936-4594}\inst{\ref{aff87}}
\and S.~Ligori\orcid{0000-0003-4172-4606}\inst{\ref{aff48}}
\and P.~B.~Lilje\orcid{0000-0003-4324-7794}\inst{\ref{aff74}}
\and V.~Lindholm\orcid{0000-0003-2317-5471}\inst{\ref{aff85},\ref{aff86}}
\and I.~Lloro\orcid{0000-0001-5966-1434}\inst{\ref{aff88}}
\and G.~Mainetti\orcid{0000-0003-2384-2377}\inst{\ref{aff89}}
\and D.~Maino\inst{\ref{aff90},\ref{aff50},\ref{aff91}}
\and E.~Maiorano\orcid{0000-0003-2593-4355}\inst{\ref{aff2}}
\and O.~Mansutti\orcid{0000-0001-5758-4658}\inst{\ref{aff7}}
\and O.~Marggraf\orcid{0000-0001-7242-3852}\inst{\ref{aff92}}
\and M.~Martinelli\orcid{0000-0002-6943-7732}\inst{\ref{aff55},\ref{aff93}}
\and N.~Martinet\orcid{0000-0003-2786-7790}\inst{\ref{aff94}}
\and F.~Marulli\orcid{0000-0002-8850-0303}\inst{\ref{aff20},\ref{aff2},\ref{aff36}}
\and R.~J.~Massey\orcid{0000-0002-6085-3780}\inst{\ref{aff95}}
\and E.~Medinaceli\orcid{0000-0002-4040-7783}\inst{\ref{aff2}}
\and S.~Mei\orcid{0000-0002-2849-559X}\inst{\ref{aff6},\ref{aff96}}
\and M.~Melchior\inst{\ref{aff97}}
\and Y.~Mellier\thanks{Deceased}\inst{\ref{aff15},\ref{aff14}}
\and M.~Meneghetti\orcid{0000-0003-1225-7084}\inst{\ref{aff2},\ref{aff36}}
\and E.~Merlin\orcid{0000-0001-6870-8900}\inst{\ref{aff55}}
\and G.~Meylan\inst{\ref{aff98}}
\and A.~Mora\orcid{0000-0002-1922-8529}\inst{\ref{aff99}}
\and M.~Moresco\orcid{0000-0002-7616-7136}\inst{\ref{aff20},\ref{aff2}}
\and L.~Moscardini\orcid{0000-0002-3473-6716}\inst{\ref{aff20},\ref{aff2},\ref{aff36}}
\and R.~Nakajima\orcid{0009-0009-1213-7040}\inst{\ref{aff92}}
\and S.-M.~Niemi\orcid{0009-0005-0247-0086}\inst{\ref{aff49}}
\and C.~Padilla\orcid{0000-0001-7951-0166}\inst{\ref{aff100}}
\and S.~Paltani\orcid{0000-0002-8108-9179}\inst{\ref{aff26}}
\and F.~Pasian\orcid{0000-0002-4869-3227}\inst{\ref{aff7}}
\and K.~Pedersen\inst{\ref{aff101}}
\and W.~J.~Percival\orcid{0000-0002-0644-5727}\inst{\ref{aff102},\ref{aff103},\ref{aff104}}
\and V.~Pettorino\orcid{0000-0002-4203-9320}\inst{\ref{aff49}}
\and S.~Pires\orcid{0000-0002-0249-2104}\inst{\ref{aff3}}
\and G.~Polenta\orcid{0000-0003-4067-9196}\inst{\ref{aff105}}
\and M.~Poncet\inst{\ref{aff106}}
\and L.~A.~Popa\inst{\ref{aff107}}
\and F.~Raison\orcid{0000-0002-7819-6918}\inst{\ref{aff38}}
\and A.~Renzi\orcid{0000-0001-9856-1970}\inst{\ref{aff108},\ref{aff69}}
\and J.~Rhodes\orcid{0000-0002-4485-8549}\inst{\ref{aff75}}
\and G.~Riccio\inst{\ref{aff43}}
\and E.~Romelli\orcid{0000-0003-3069-9222}\inst{\ref{aff7}}
\and M.~Roncarelli\orcid{0000-0001-9587-7822}\inst{\ref{aff2}}
\and R.~Saglia\orcid{0000-0003-0378-7032}\inst{\ref{aff39},\ref{aff38}}
\and Z.~Sakr\orcid{0000-0002-4823-3757}\inst{\ref{aff109},\ref{aff110},\ref{aff111}}
\and D.~Sapone\orcid{0000-0001-7089-4503}\inst{\ref{aff112}}
\and B.~Sartoris\orcid{0000-0003-1337-5269}\inst{\ref{aff39},\ref{aff7}}
\and P.~Schneider\orcid{0000-0001-8561-2679}\inst{\ref{aff92}}
\and T.~Schrabback\orcid{0000-0002-6987-7834}\inst{\ref{aff113}}
\and A.~Secroun\orcid{0000-0003-0505-3710}\inst{\ref{aff70}}
\and G.~Seidel\orcid{0000-0003-2907-353X}\inst{\ref{aff80}}
\and S.~Serrano\orcid{0000-0002-0211-2861}\inst{\ref{aff114},\ref{aff115},\ref{aff116}}
\and P.~Simon\inst{\ref{aff92}}
\and C.~Sirignano\orcid{0000-0002-0995-7146}\inst{\ref{aff108},\ref{aff69}}
\and G.~Sirri\orcid{0000-0003-2626-2853}\inst{\ref{aff36}}
\and J.~Skottfelt\orcid{0000-0003-1310-8283}\inst{\ref{aff117}}
\and L.~Stanco\orcid{0000-0002-9706-5104}\inst{\ref{aff69}}
\and J.~Steinwagner\orcid{0000-0001-7443-1047}\inst{\ref{aff38}}
\and P.~Tallada-Cresp\'{i}\orcid{0000-0002-1336-8328}\inst{\ref{aff51},\ref{aff52}}
\and A.~N.~Taylor\inst{\ref{aff60}}
\and H.~I.~Teplitz\orcid{0000-0002-7064-5424}\inst{\ref{aff118}}
\and I.~Tereno\orcid{0000-0002-4537-6218}\inst{\ref{aff67},\ref{aff119}}
\and N.~Tessore\orcid{0000-0002-9696-7931}\inst{\ref{aff82}}
\and S.~Toft\orcid{0000-0003-3631-7176}\inst{\ref{aff27},\ref{aff28}}
\and R.~Toledo-Moreo\orcid{0000-0002-2997-4859}\inst{\ref{aff120}}
\and F.~Torradeflot\orcid{0000-0003-1160-1517}\inst{\ref{aff52},\ref{aff51}}
\and I.~Tutusaus\orcid{0000-0002-3199-0399}\inst{\ref{aff116},\ref{aff114},\ref{aff110}}
\and L.~Valenziano\orcid{0000-0002-1170-0104}\inst{\ref{aff2},\ref{aff71}}
\and J.~Valiviita\orcid{0000-0001-6225-3693}\inst{\ref{aff85},\ref{aff86}}
\and T.~Vassallo\orcid{0000-0001-6512-6358}\inst{\ref{aff7},\ref{aff72}}
\and G.~Verdoes~Kleijn\orcid{0000-0001-5803-2580}\inst{\ref{aff5}}
\and A.~Veropalumbo\orcid{0000-0003-2387-1194}\inst{\ref{aff10},\ref{aff41},\ref{aff40}}
\and Y.~Wang\orcid{0000-0002-4749-2984}\inst{\ref{aff118}}
\and J.~Weller\orcid{0000-0002-8282-2010}\inst{\ref{aff39},\ref{aff38}}
\and A.~Zacchei\orcid{0000-0003-0396-1192}\inst{\ref{aff7},\ref{aff33}}
\and G.~Zamorani\orcid{0000-0002-2318-301X}\inst{\ref{aff2}}
\and I.~A.~Zinchenko\orcid{0000-0002-2944-2449}\inst{\ref{aff121}}
\and E.~Zucca\orcid{0000-0002-5845-8132}\inst{\ref{aff2}}
\and V.~Allevato\orcid{0000-0001-7232-5152}\inst{\ref{aff43}}
\and M.~Ballardini\orcid{0000-0003-4481-3559}\inst{\ref{aff122},\ref{aff123},\ref{aff2}}
\and E.~Bozzo\orcid{0000-0002-8201-1525}\inst{\ref{aff26}}
\and C.~Burigana\orcid{0000-0002-3005-5796}\inst{\ref{aff124},\ref{aff71}}
\and R.~Cabanac\orcid{0000-0001-6679-2600}\inst{\ref{aff110}}
\and M.~Calabrese\orcid{0000-0002-2637-2422}\inst{\ref{aff125},\ref{aff50}}
\and A.~Cappi\inst{\ref{aff2},\ref{aff126}}
\and D.~Di~Ferdinando\inst{\ref{aff36}}
\and J.~A.~Escartin~Vigo\inst{\ref{aff38}}
\and W.~G.~Hartley\inst{\ref{aff26}}
\and M.~Huertas-Company\orcid{0000-0002-1416-8483}\inst{\ref{aff59},\ref{aff127},\ref{aff128},\ref{aff129}}
\and J.~Mart\'{i}n-Fleitas\orcid{0000-0002-8594-569X}\inst{\ref{aff130}}
\and S.~Matthew\orcid{0000-0001-8448-1697}\inst{\ref{aff60}}
\and N.~Mauri\orcid{0000-0001-8196-1548}\inst{\ref{aff58},\ref{aff36}}
\and R.~B.~Metcalf\orcid{0000-0003-3167-2574}\inst{\ref{aff20},\ref{aff2}}
\and A.~Pezzotta\orcid{0000-0003-0726-2268}\inst{\ref{aff10}}
\and M.~P\"ontinen\orcid{0000-0001-5442-2530}\inst{\ref{aff85}}
\and I.~Risso\orcid{0000-0003-2525-7761}\inst{\ref{aff10},\ref{aff41}}
\and V.~Scottez\orcid{0009-0008-3864-940X}\inst{\ref{aff15},\ref{aff131}}
\and M.~Sereno\orcid{0000-0003-0302-0325}\inst{\ref{aff2},\ref{aff36}}
\and M.~Tenti\orcid{0000-0002-4254-5901}\inst{\ref{aff36}}
\and M.~Viel\orcid{0000-0002-2642-5707}\inst{\ref{aff33},\ref{aff7},\ref{aff35},\ref{aff34},\ref{aff132}}
\and M.~Wiesmann\orcid{0009-0000-8199-5860}\inst{\ref{aff74}}
\and Y.~Akrami\orcid{0000-0002-2407-7956}\inst{\ref{aff133},\ref{aff134}}
\and I.~T.~Andika\orcid{0000-0001-6102-9526}\inst{\ref{aff135},\ref{aff136}}
\and S.~Anselmi\orcid{0000-0002-3579-9583}\inst{\ref{aff69},\ref{aff108},\ref{aff137}}
\and M.~Archidiacono\orcid{0000-0003-4952-9012}\inst{\ref{aff90},\ref{aff91}}
\and F.~Atrio-Barandela\orcid{0000-0002-2130-2513}\inst{\ref{aff138}}
\and D.~Bertacca\orcid{0000-0002-2490-7139}\inst{\ref{aff108},\ref{aff37},\ref{aff69}}
\and M.~Bethermin\orcid{0000-0002-3915-2015}\inst{\ref{aff139}}
\and L.~Bisigello\orcid{0000-0003-0492-4924}\inst{\ref{aff37}}
\and A.~Blanchard\orcid{0000-0001-8555-9003}\inst{\ref{aff110}}
\and L.~Blot\orcid{0000-0002-9622-7167}\inst{\ref{aff140},\ref{aff87}}
\and H.~B\"ohringer\orcid{0000-0001-8241-4204}\inst{\ref{aff38},\ref{aff141},\ref{aff142}}
\and M.~Bonici\orcid{0000-0002-8430-126X}\inst{\ref{aff102},\ref{aff50}}
\and S.~Borgani\orcid{0000-0001-6151-6439}\inst{\ref{aff24},\ref{aff33},\ref{aff7},\ref{aff34},\ref{aff132}}
\and M.~L.~Brown\orcid{0000-0002-0370-8077}\inst{\ref{aff143}}
\and S.~Bruton\orcid{0000-0002-6503-5218}\inst{\ref{aff144}}
\and A.~Calabro\orcid{0000-0003-2536-1614}\inst{\ref{aff55}}
\and B.~Camacho~Quevedo\orcid{0000-0002-8789-4232}\inst{\ref{aff33},\ref{aff35},\ref{aff7}}
\and F.~Caro\inst{\ref{aff55}}
\and C.~S.~Carvalho\inst{\ref{aff119}}
\and T.~Castro\orcid{0000-0002-6292-3228}\inst{\ref{aff7},\ref{aff34},\ref{aff33},\ref{aff132}}
\and F.~Cogato\orcid{0000-0003-4632-6113}\inst{\ref{aff20},\ref{aff2}}
\and S.~Conseil\orcid{0000-0002-3657-4191}\inst{\ref{aff62}}
\and T.~Contini\orcid{0000-0003-0275-938X}\inst{\ref{aff110}}
\and A.~R.~Cooray\orcid{0000-0002-3892-0190}\inst{\ref{aff145}}
\and O.~Cucciati\orcid{0000-0002-9336-7551}\inst{\ref{aff2}}
\and G.~Desprez\orcid{0000-0001-8325-1742}\inst{\ref{aff5}}
\and A.~D\'iaz-S\'anchez\orcid{0000-0003-0748-4768}\inst{\ref{aff146}}
\and S.~Di~Domizio\orcid{0000-0003-2863-5895}\inst{\ref{aff40},\ref{aff41}}
\and J.~M.~Diego\orcid{0000-0001-9065-3926}\inst{\ref{aff147}}
\and P.~Dimauro\orcid{0000-0001-7399-2854}\inst{\ref{aff148},\ref{aff55}}
\and P.-A.~Duc\orcid{0000-0003-3343-6284}\inst{\ref{aff139}}
\and M.~Y.~Elkhashab\orcid{0000-0001-9306-2603}\inst{\ref{aff7},\ref{aff34},\ref{aff24},\ref{aff33}}
\and Y.~Fang\orcid{0000-0002-0334-6950}\inst{\ref{aff39}}
\and A.~Finoguenov\orcid{0000-0002-4606-5403}\inst{\ref{aff85}}
\and A.~Fontana\orcid{0000-0003-3820-2823}\inst{\ref{aff55}}
\and F.~Fontanot\orcid{0000-0003-4744-0188}\inst{\ref{aff7},\ref{aff33}}
\and A.~Franco\orcid{0000-0002-4761-366X}\inst{\ref{aff149},\ref{aff150},\ref{aff151}}
\and K.~Ganga\orcid{0000-0001-8159-8208}\inst{\ref{aff6}}
\and J.~Garc\'ia-Bellido\orcid{0000-0002-9370-8360}\inst{\ref{aff133}}
\and T.~Gasparetto\orcid{0000-0002-7913-4866}\inst{\ref{aff55}}
\and V.~Gautard\inst{\ref{aff1}}
\and R.~Gavazzi\orcid{0000-0002-5540-6935}\inst{\ref{aff94},\ref{aff14}}
\and E.~Gaztanaga\orcid{0000-0001-9632-0815}\inst{\ref{aff116},\ref{aff114},\ref{aff152}}
\and F.~Giacomini\orcid{0000-0002-3129-2814}\inst{\ref{aff36}}
\and F.~Gianotti\orcid{0000-0003-4666-119X}\inst{\ref{aff2}}
\and A.~H.~Gonzalez\orcid{0000-0002-0933-8601}\inst{\ref{aff153}}
\and G.~Gozaliasl\orcid{0000-0002-0236-919X}\inst{\ref{aff154},\ref{aff85}}
\and M.~Guidi\orcid{0000-0001-9408-1101}\inst{\ref{aff4},\ref{aff2}}
\and C.~M.~Gutierrez\orcid{0000-0001-7854-783X}\inst{\ref{aff155}}
\and A.~Hall\orcid{0000-0002-3139-8651}\inst{\ref{aff60}}
\and S.~Hemmati\orcid{0000-0003-2226-5395}\inst{\ref{aff156}}
\and H.~Hildebrandt\orcid{0000-0002-9814-3338}\inst{\ref{aff157}}
\and J.~Hjorth\orcid{0000-0002-4571-2306}\inst{\ref{aff101}}
\and J.~J.~E.~Kajava\orcid{0000-0002-3010-8333}\inst{\ref{aff158},\ref{aff159}}
\and Y.~Kang\orcid{0009-0000-8588-7250}\inst{\ref{aff26}}
\and V.~Kansal\orcid{0000-0002-4008-6078}\inst{\ref{aff160},\ref{aff161}}
\and D.~Karagiannis\orcid{0000-0002-4927-0816}\inst{\ref{aff122},\ref{aff162}}
\and K.~Kiiveri\inst{\ref{aff83}}
\and J.~Kim\orcid{0000-0003-2776-2761}\inst{\ref{aff19}}
\and C.~C.~Kirkpatrick\inst{\ref{aff83}}
\and S.~Kruk\orcid{0000-0001-8010-8879}\inst{\ref{aff31}}
\and L.~Legrand\orcid{0000-0003-0610-5252}\inst{\ref{aff163},\ref{aff164}}
\and M.~Lembo\orcid{0000-0002-5271-5070}\inst{\ref{aff14},\ref{aff122},\ref{aff123}}
\and F.~Lepori\orcid{0009-0000-5061-7138}\inst{\ref{aff165}}
\and G.~Leroy\orcid{0009-0004-2523-4425}\inst{\ref{aff166},\ref{aff95}}
\and G.~F.~Lesci\orcid{0000-0002-4607-2830}\inst{\ref{aff20},\ref{aff2}}
\and J.~Lesgourgues\orcid{0000-0001-7627-353X}\inst{\ref{aff53}}
\and L.~Leuzzi\orcid{0009-0006-4479-7017}\inst{\ref{aff2}}
\and T.~I.~Liaudat\orcid{0000-0002-9104-314X}\inst{\ref{aff167}}
\and A.~Loureiro\orcid{0000-0002-4371-0876}\inst{\ref{aff168},\ref{aff169}}
\and J.~Macias-Perez\orcid{0000-0002-5385-2763}\inst{\ref{aff170}}
\and E.~A.~Magnier\orcid{0000-0002-7965-2815}\inst{\ref{aff57}}
\and F.~Mannucci\orcid{0000-0002-4803-2381}\inst{\ref{aff171}}
\and R.~Maoli\orcid{0000-0002-6065-3025}\inst{\ref{aff172},\ref{aff55}}
\and C.~J.~A.~P.~Martins\orcid{0000-0002-4886-9261}\inst{\ref{aff173},\ref{aff11}}
\and L.~Maurin\orcid{0000-0002-8406-0857}\inst{\ref{aff18}}
\and M.~Miluzio\inst{\ref{aff31},\ref{aff174}}
\and P.~Monaco\orcid{0000-0003-2083-7564}\inst{\ref{aff24},\ref{aff7},\ref{aff34},\ref{aff33},\ref{aff132}}
\and C.~Moretti\orcid{0000-0003-3314-8936}\inst{\ref{aff7},\ref{aff33},\ref{aff34},\ref{aff35}}
\and G.~Morgante\inst{\ref{aff2}}
\and K.~Naidoo\orcid{0000-0002-9182-1802}\inst{\ref{aff152},\ref{aff82}}
\and A.~Navarro-Alsina\orcid{0000-0002-3173-2592}\inst{\ref{aff92}}
\and S.~Nesseris\orcid{0000-0002-0567-0324}\inst{\ref{aff133}}
\and D.~Paoletti\orcid{0000-0003-4761-6147}\inst{\ref{aff2},\ref{aff71}}
\and F.~Passalacqua\orcid{0000-0002-8606-4093}\inst{\ref{aff108},\ref{aff69}}
\and K.~Paterson\orcid{0000-0001-8340-3486}\inst{\ref{aff80}}
\and L.~Patrizii\inst{\ref{aff36}}
\and A.~Pisani\orcid{0000-0002-6146-4437}\inst{\ref{aff70}}
\and D.~Potter\orcid{0000-0002-0757-5195}\inst{\ref{aff165}}
\and M.~Radovich\orcid{0000-0002-3585-866X}\inst{\ref{aff37}}
\and G.~Rodighiero\orcid{0000-0002-9415-2296}\inst{\ref{aff108},\ref{aff37}}
\and S.~Sacquegna\orcid{0000-0002-8433-6630}\inst{\ref{aff175},\ref{aff150},\ref{aff149}}
\and M.~Sahl\'en\orcid{0000-0003-0973-4804}\inst{\ref{aff176}}
\and D.~B.~Sanders\orcid{0000-0002-1233-9998}\inst{\ref{aff57}}
\and E.~Sarpa\orcid{0000-0002-1256-655X}\inst{\ref{aff35},\ref{aff132},\ref{aff34}}
\and C.~Scarlata\orcid{0000-0002-9136-8876}\inst{\ref{aff177}}
\and A.~Schneider\orcid{0000-0001-7055-8104}\inst{\ref{aff165}}
\and M.~Schultheis\inst{\ref{aff126}}
\and D.~Sciotti\orcid{0009-0008-4519-2620}\inst{\ref{aff55},\ref{aff93}}
\and E.~Sellentin\inst{\ref{aff178},\ref{aff29}}
\and L.~C.~Smith\orcid{0000-0002-3259-2771}\inst{\ref{aff179}}
\and S.~A.~Stanford\orcid{0000-0003-0122-0841}\inst{\ref{aff180}}
\and K.~Tanidis\orcid{0000-0001-9843-5130}\inst{\ref{aff19}}
\and G.~Testera\inst{\ref{aff41}}
\and R.~Teyssier\orcid{0000-0001-7689-0933}\inst{\ref{aff181}}
\and S.~Tosi\orcid{0000-0002-7275-9193}\inst{\ref{aff40},\ref{aff41},\ref{aff10}}
\and A.~Troja\orcid{0000-0003-0239-4595}\inst{\ref{aff108},\ref{aff69}}
\and M.~Tucci\inst{\ref{aff26}}
\and C.~Valieri\inst{\ref{aff36}}
\and A.~Venhola\orcid{0000-0001-6071-4564}\inst{\ref{aff182}}
\and D.~Vergani\orcid{0000-0003-0898-2216}\inst{\ref{aff2}}
\and G.~Verza\orcid{0000-0002-1886-8348}\inst{\ref{aff183}}
\and P.~Vielzeuf\orcid{0000-0003-2035-9339}\inst{\ref{aff70}}
\and N.~A.~Walton\orcid{0000-0003-3983-8778}\inst{\ref{aff179}}}

\institute{CEA Saclay, DFR/IRFU, Service d'Astrophysique, Bat. 709, 91191 Gif-sur-Yvette, France\label{aff1}
\and
INAF-Osservatorio di Astrofisica e Scienza dello Spazio di Bologna, Via Piero Gobetti 93/3, 40129 Bologna, Italy\label{aff2}
\and
Universit\'e Paris-Saclay, Universit\'e Paris Cit\'e, CEA, CNRS, AIM, 91191, Gif-sur-Yvette, France\label{aff3}
\and
Dipartimento di Fisica e Astronomia, Universit\`a di Bologna, Via Gobetti 93/2, 40129 Bologna, Italy\label{aff4}
\and
Kapteyn Astronomical Institute, University of Groningen, PO Box 800, 9700 AV Groningen, The Netherlands\label{aff5}
\and
Universit\'e Paris Cit\'e, CNRS, Astroparticule et Cosmologie, 75013 Paris, France\label{aff6}
\and
INAF-Osservatorio Astronomico di Trieste, Via G. B. Tiepolo 11, 34143 Trieste, Italy\label{aff7}
\and
Institute de Physique du Globe de Paris, 1 Rue Jussieu, 75005, Paris\label{aff8}
\and
Dipartimento di Fisica ``G. Occhialini", Universit\`a degli Studi di Milano Bicocca, Piazza della Scienza 3, 20126 Milano, Italy\label{aff9}
\and
INAF-Osservatorio Astronomico di Brera, Via Brera 28, 20122 Milano, Italy\label{aff10}
\and
Instituto de Astrof\'isica e Ci\^encias do Espa\c{c}o, Universidade do Porto, CAUP, Rua das Estrelas, PT4150-762 Porto, Portugal\label{aff11}
\and
Departamento de F\'{\i}sica e Astronomia, Faculdade de Ci\^encias, Universidade do Porto, Rua do Campo Alegre 687, PT4169-007 Porto, Portugal\label{aff12}
\and
INAF-Istituto di Astrofisica e Planetologia Spaziali, via del Fosso del Cavaliere, 100, 00100 Roma, Italy\label{aff13}
\and
Institut d'Astrophysique de Paris, UMR 7095, CNRS, and Sorbonne Universit\'e, 98 bis boulevard Arago, 75014 Paris, France\label{aff14}
\and
Institut d'Astrophysique de Paris, 98bis Boulevard Arago, 75014, Paris, France\label{aff15}
\and
Centre de Recherche Astrophysique de Lyon, UMR5574, CNRS, Universit\'e Claude Bernard Lyon 1, ENS de Lyon, 69230, Saint-Genis-Laval, France\label{aff16}
\and
Univ. Lille, CNRS, Centrale Lille, UMR 9189 CRIStAL, 59000 Lille, France\label{aff17}
\and
Universit\'e Paris-Saclay, CNRS, Institut d'astrophysique spatiale, 91405, Orsay, France\label{aff18}
\and
Department of Physics, Oxford University, Keble Road, Oxford OX1 3RH, UK\label{aff19}
\and
Dipartimento di Fisica e Astronomia "Augusto Righi" - Alma Mater Studiorum Universit\`a di Bologna, via Piero Gobetti 93/2, 40129 Bologna, Italy\label{aff20}
\and
School of Physics \& Astronomy, University of Southampton, Highfield Campus, Southampton SO17 1BJ, UK\label{aff21}
\and
Sterrenkundig Observatorium, Universiteit Gent, Krijgslaan 281 S9, 9000 Gent, Belgium\label{aff22}
\and
Center for Astronomy and Astrophysics and Department of Physics, Fudan University, Shanghai 200438, People's Republic of China\label{aff23}
\and
Dipartimento di Fisica - Sezione di Astronomia, Universit\`a di Trieste, Via Tiepolo 11, 34131 Trieste, Italy\label{aff24}
\and
Institute of Science and Technology Austria (ISTA), Am Campus 1, 3400 Klosterneuburg, Austria\label{aff25}
\and
Department of Astronomy, University of Geneva, ch. d'Ecogia 16, 1290 Versoix, Switzerland\label{aff26}
\and
Cosmic Dawn Center (DAWN)\label{aff27}
\and
Niels Bohr Institute, University of Copenhagen, Jagtvej 128, 2200 Copenhagen, Denmark\label{aff28}
\and
Leiden Observatory, Leiden University, Einsteinweg 55, 2333 CC Leiden, The Netherlands\label{aff29}
\and
Department of Physics and Astronomy, University of British Columbia, Vancouver, BC V6T 1Z1, Canada\label{aff30}
\and
ESAC/ESA, Camino Bajo del Castillo, s/n., Urb. Villafranca del Castillo, 28692 Villanueva de la Ca\~nada, Madrid, Spain\label{aff31}
\and
School of Mathematics and Physics, University of Surrey, Guildford, Surrey, GU2 7XH, UK\label{aff32}
\and
IFPU, Institute for Fundamental Physics of the Universe, via Beirut 2, 34151 Trieste, Italy\label{aff33}
\and
INFN, Sezione di Trieste, Via Valerio 2, 34127 Trieste TS, Italy\label{aff34}
\and
SISSA, International School for Advanced Studies, Via Bonomea 265, 34136 Trieste TS, Italy\label{aff35}
\and
INFN-Sezione di Bologna, Viale Berti Pichat 6/2, 40127 Bologna, Italy\label{aff36}
\and
INAF-Osservatorio Astronomico di Padova, Via dell'Osservatorio 5, 35122 Padova, Italy\label{aff37}
\and
Max Planck Institute for Extraterrestrial Physics, Giessenbachstr. 1, 85748 Garching, Germany\label{aff38}
\and
Universit\"ats-Sternwarte M\"unchen, Fakult\"at f\"ur Physik, Ludwig-Maximilians-Universit\"at M\"unchen, Scheinerstrasse 1, 81679 M\"unchen, Germany\label{aff39}
\and
Dipartimento di Fisica, Universit\`a di Genova, Via Dodecaneso 33, 16146, Genova, Italy\label{aff40}
\and
INFN-Sezione di Genova, Via Dodecaneso 33, 16146, Genova, Italy\label{aff41}
\and
Department of Physics "E. Pancini", University Federico II, Via Cinthia 6, 80126, Napoli, Italy\label{aff42}
\and
INAF-Osservatorio Astronomico di Capodimonte, Via Moiariello 16, 80131 Napoli, Italy\label{aff43}
\and
Faculdade de Ci\^encias da Universidade do Porto, Rua do Campo de Alegre, 4150-007 Porto, Portugal\label{aff44}
\and
European Southern Observatory, Karl-Schwarzschild-Str.~2, 85748 Garching, Germany\label{aff45}
\and
Dipartimento di Fisica, Universit\`a degli Studi di Torino, Via P. Giuria 1, 10125 Torino, Italy\label{aff46}
\and
INFN-Sezione di Torino, Via P. Giuria 1, 10125 Torino, Italy\label{aff47}
\and
INAF-Osservatorio Astrofisico di Torino, Via Osservatorio 20, 10025 Pino Torinese (TO), Italy\label{aff48}
\and
European Space Agency/ESTEC, Keplerlaan 1, 2201 AZ Noordwijk, The Netherlands\label{aff49}
\and
INAF-IASF Milano, Via Alfonso Corti 12, 20133 Milano, Italy\label{aff50}
\and
Centro de Investigaciones Energ\'eticas, Medioambientales y Tecnol\'ogicas (CIEMAT), Avenida Complutense 40, 28040 Madrid, Spain\label{aff51}
\and
Port d'Informaci\'{o} Cient\'{i}fica, Campus UAB, C. Albareda s/n, 08193 Bellaterra (Barcelona), Spain\label{aff52}
\and
Institute for Theoretical Particle Physics and Cosmology (TTK), RWTH Aachen University, 52056 Aachen, Germany\label{aff53}
\and
Deutsches Zentrum f\"ur Luft- und Raumfahrt e. V. (DLR), Linder H\"ohe, 51147 K\"oln, Germany\label{aff54}
\and
INAF-Osservatorio Astronomico di Roma, Via Frascati 33, 00078 Monteporzio Catone, Italy\label{aff55}
\and
INFN section of Naples, Via Cinthia 6, 80126, Napoli, Italy\label{aff56}
\and
Institute for Astronomy, University of Hawaii, 2680 Woodlawn Drive, Honolulu, HI 96822, USA\label{aff57}
\and
Dipartimento di Fisica e Astronomia "Augusto Righi" - Alma Mater Studiorum Universit\`a di Bologna, Viale Berti Pichat 6/2, 40127 Bologna, Italy\label{aff58}
\and
Instituto de Astrof\'{\i}sica de Canarias, V\'{\i}a L\'actea, 38205 La Laguna, Tenerife, Spain\label{aff59}
\and
Institute for Astronomy, University of Edinburgh, Royal Observatory, Blackford Hill, Edinburgh EH9 3HJ, UK\label{aff60}
\and
European Space Agency/ESRIN, Largo Galileo Galilei 1, 00044 Frascati, Roma, Italy\label{aff61}
\and
Universit\'e Claude Bernard Lyon 1, CNRS/IN2P3, IP2I Lyon, UMR 5822, Villeurbanne, F-69100, France\label{aff62}
\and
Institut de Ci\`{e}ncies del Cosmos (ICCUB), Universitat de Barcelona (IEEC-UB), Mart\'{i} i Franqu\`{e}s 1, 08028 Barcelona, Spain\label{aff63}
\and
Instituci\'o Catalana de Recerca i Estudis Avan\c{c}ats (ICREA), Passeig de Llu\'{\i}s Companys 23, 08010 Barcelona, Spain\label{aff64}
\and
UCB Lyon 1, CNRS/IN2P3, IUF, IP2I Lyon, 4 rue Enrico Fermi, 69622 Villeurbanne, France\label{aff65}
\and
Mullard Space Science Laboratory, University College London, Holmbury St Mary, Dorking, Surrey RH5 6NT, UK\label{aff66}
\and
Departamento de F\'isica, Faculdade de Ci\^encias, Universidade de Lisboa, Edif\'icio C8, Campo Grande, PT1749-016 Lisboa, Portugal\label{aff67}
\and
Instituto de Astrof\'isica e Ci\^encias do Espa\c{c}o, Faculdade de Ci\^encias, Universidade de Lisboa, Campo Grande, 1749-016 Lisboa, Portugal\label{aff68}
\and
INFN-Padova, Via Marzolo 8, 35131 Padova, Italy\label{aff69}
\and
Aix-Marseille Universit\'e, CNRS/IN2P3, CPPM, Marseille, France\label{aff70}
\and
INFN-Bologna, Via Irnerio 46, 40126 Bologna, Italy\label{aff71}
\and
University Observatory, LMU Faculty of Physics, Scheinerstrasse 1, 81679 Munich, Germany\label{aff72}
\and
Herzberg Astronomy and Astrophysics Research Centre, 5071 W. Saanich Rd. Victoria, BC, V9E 2E7, Canada\label{aff73}
\and
Institute of Theoretical Astrophysics, University of Oslo, P.O. Box 1029 Blindern, 0315 Oslo, Norway\label{aff74}
\and
Jet Propulsion Laboratory, California Institute of Technology, 4800 Oak Grove Drive, Pasadena, CA, 91109, USA\label{aff75}
\and
Department of Physics, Lancaster University, Lancaster, LA1 4YB, UK\label{aff76}
\and
Felix Hormuth Engineering, Goethestr. 17, 69181 Leimen, Germany\label{aff77}
\and
Technical University of Denmark, Elektrovej 327, 2800 Kgs. Lyngby, Denmark\label{aff78}
\and
Cosmic Dawn Center (DAWN), Denmark\label{aff79}
\and
Max-Planck-Institut f\"ur Astronomie, K\"onigstuhl 17, 69117 Heidelberg, Germany\label{aff80}
\and
NASA Goddard Space Flight Center, Greenbelt, MD 20771, USA\label{aff81}
\and
Department of Physics and Astronomy, University College London, Gower Street, London WC1E 6BT, UK\label{aff82}
\and
Department of Physics and Helsinki Institute of Physics, Gustaf H\"allstr\"omin katu 2, University of Helsinki, 00014 Helsinki, Finland\label{aff83}
\and
Universit\'e de Gen\`eve, D\'epartement de Physique Th\'eorique and Centre for Astroparticle Physics, 24 quai Ernest-Ansermet, CH-1211 Gen\`eve 4, Switzerland\label{aff84}
\and
Department of Physics, P.O. Box 64, University of Helsinki, 00014 Helsinki, Finland\label{aff85}
\and
Helsinki Institute of Physics, Gustaf H{\"a}llstr{\"o}min katu 2, University of Helsinki, 00014 Helsinki, Finland\label{aff86}
\and
Laboratoire d'etude de l'Univers et des phenomenes eXtremes, Observatoire de Paris, Universit\'e PSL, Sorbonne Universit\'e, CNRS, 92190 Meudon, France\label{aff87}
\and
SKAO, Jodrell Bank, Lower Withington, Macclesfield SK11 9FT, UK\label{aff88}
\and
Centre de Calcul de l'IN2P3/CNRS, 21 avenue Pierre de Coubertin 69627 Villeurbanne Cedex, France\label{aff89}
\and
Dipartimento di Fisica "Aldo Pontremoli", Universit\`a degli Studi di Milano, Via Celoria 16, 20133 Milano, Italy\label{aff90}
\and
INFN-Sezione di Milano, Via Celoria 16, 20133 Milano, Italy\label{aff91}
\and
Universit\"at Bonn, Argelander-Institut f\"ur Astronomie, Auf dem H\"ugel 71, 53121 Bonn, Germany\label{aff92}
\and
INFN-Sezione di Roma, Piazzale Aldo Moro, 2 - c/o Dipartimento di Fisica, Edificio G. Marconi, 00185 Roma, Italy\label{aff93}
\and
Aix-Marseille Universit\'e, CNRS, CNES, LAM, Marseille, France\label{aff94}
\and
Department of Physics, Institute for Computational Cosmology, Durham University, South Road, Durham, DH1 3LE, UK\label{aff95}
\and
CNRS-UCB International Research Laboratory, Centre Pierre Bin\'etruy, IRL2007, CPB-IN2P3, Berkeley, USA\label{aff96}
\and
University of Applied Sciences and Arts of Northwestern Switzerland, School of Engineering, 5210 Windisch, Switzerland\label{aff97}
\and
Institute of Physics, Laboratory of Astrophysics, Ecole Polytechnique F\'ed\'erale de Lausanne (EPFL), Observatoire de Sauverny, 1290 Versoix, Switzerland\label{aff98}
\and
Telespazio UK S.L. for European Space Agency (ESA), Camino bajo del Castillo, s/n, Urbanizacion Villafranca del Castillo, Villanueva de la Ca\~nada, 28692 Madrid, Spain\label{aff99}
\and
Institut de F\'{i}sica d'Altes Energies (IFAE), The Barcelona Institute of Science and Technology, Campus UAB, 08193 Bellaterra (Barcelona), Spain\label{aff100}
\and
DARK, Niels Bohr Institute, University of Copenhagen, Jagtvej 155, 2200 Copenhagen, Denmark\label{aff101}
\and
Waterloo Centre for Astrophysics, University of Waterloo, Waterloo, Ontario N2L 3G1, Canada\label{aff102}
\and
Department of Physics and Astronomy, University of Waterloo, Waterloo, Ontario N2L 3G1, Canada\label{aff103}
\and
Perimeter Institute for Theoretical Physics, Waterloo, Ontario N2L 2Y5, Canada\label{aff104}
\and
Space Science Data Center, Italian Space Agency, via del Politecnico snc, 00133 Roma, Italy\label{aff105}
\and
Centre National d'Etudes Spatiales -- Centre spatial de Toulouse, 18 avenue Edouard Belin, 31401 Toulouse Cedex 9, France\label{aff106}
\and
Institute of Space Science, Str. Atomistilor, nr. 409 M\u{a}gurele, Ilfov, 077125, Romania\label{aff107}
\and
Dipartimento di Fisica e Astronomia "G. Galilei", Universit\`a di Padova, Via Marzolo 8, 35131 Padova, Italy\label{aff108}
\and
Institut f\"ur Theoretische Physik, University of Heidelberg, Philosophenweg 16, 69120 Heidelberg, Germany\label{aff109}
\and
Institut de Recherche en Astrophysique et Plan\'etologie (IRAP), Universit\'e de Toulouse, CNRS, UPS, CNES, 14 Av. Edouard Belin, 31400 Toulouse, France\label{aff110}
\and
Universit\'e St Joseph; Faculty of Sciences, Beirut, Lebanon\label{aff111}
\and
Departamento de F\'isica, FCFM, Universidad de Chile, Blanco Encalada 2008, Santiago, Chile\label{aff112}
\and
Universit\"at Innsbruck, Institut f\"ur Astro- und Teilchenphysik, Technikerstr. 25/8, 6020 Innsbruck, Austria\label{aff113}
\and
Institut d'Estudis Espacials de Catalunya (IEEC),  Edifici RDIT, Campus UPC, 08860 Castelldefels, Barcelona, Spain\label{aff114}
\and
Satlantis, University Science Park, Sede Bld 48940, Leioa-Bilbao, Spain\label{aff115}
\and
Institute of Space Sciences (ICE, CSIC), Campus UAB, Carrer de Can Magrans, s/n, 08193 Barcelona, Spain\label{aff116}
\and
Centre for Electronic Imaging, Open University, Walton Hall, Milton Keynes, MK7~6AA, UK\label{aff117}
\and
Infrared Processing and Analysis Center, California Institute of Technology, Pasadena, CA 91125, USA\label{aff118}
\and
Instituto de Astrof\'isica e Ci\^encias do Espa\c{c}o, Faculdade de Ci\^encias, Universidade de Lisboa, Tapada da Ajuda, 1349-018 Lisboa, Portugal\label{aff119}
\and
Universidad Polit\'ecnica de Cartagena, Departamento de Electr\'onica y Tecnolog\'ia de Computadoras,  Plaza del Hospital 1, 30202 Cartagena, Spain\label{aff120}
\and
Astronomisches Rechen-Institut, Zentrum f\"ur Astronomie der Universit\"at Heidelberg, M\"onchhofstr. 12-14, 69120 Heidelberg, Germany\label{aff121}
\and
Dipartimento di Fisica e Scienze della Terra, Universit\`a degli Studi di Ferrara, Via Giuseppe Saragat 1, 44122 Ferrara, Italy\label{aff122}
\and
Istituto Nazionale di Fisica Nucleare, Sezione di Ferrara, Via Giuseppe Saragat 1, 44122 Ferrara, Italy\label{aff123}
\and
INAF, Istituto di Radioastronomia, Via Piero Gobetti 101, 40129 Bologna, Italy\label{aff124}
\and
Astronomical Observatory of the Autonomous Region of the Aosta Valley (OAVdA), Loc. Lignan 39, I-11020, Nus (Aosta Valley), Italy\label{aff125}
\and
Universit\'e C\^{o}te d'Azur, Observatoire de la C\^{o}te d'Azur, CNRS, Laboratoire Lagrange, Bd de l'Observatoire, CS 34229, 06304 Nice cedex 4, France\label{aff126}
\and
Instituto de Astrof\'isica de Canarias (IAC); Departamento de Astrof\'isica, Universidad de La Laguna (ULL), 38200, La Laguna, Tenerife, Spain\label{aff127}
\and
Universit\'e PSL, Observatoire de Paris, Sorbonne Universit\'e, CNRS, LERMA, 75014, Paris, France\label{aff128}
\and
Universit\'e Paris-Cit\'e, 5 Rue Thomas Mann, 75013, Paris, France\label{aff129}
\and
Aurora Technology for European Space Agency (ESA), Camino bajo del Castillo, s/n, Urbanizacion Villafranca del Castillo, Villanueva de la Ca\~nada, 28692 Madrid, Spain\label{aff130}
\and
ICL, Junia, Universit\'e Catholique de Lille, LITL, 59000 Lille, France\label{aff131}
\and
ICSC - Centro Nazionale di Ricerca in High Performance Computing, Big Data e Quantum Computing, Via Magnanelli 2, Bologna, Italy\label{aff132}
\and
Instituto de F\'isica Te\'orica UAM-CSIC, Campus de Cantoblanco, 28049 Madrid, Spain\label{aff133}
\and
CERCA/ISO, Department of Physics, Case Western Reserve University, 10900 Euclid Avenue, Cleveland, OH 44106, USA\label{aff134}
\and
Technical University of Munich, TUM School of Natural Sciences, Physics Department, James-Franck-Str.~1, 85748 Garching, Germany\label{aff135}
\and
Max-Planck-Institut f\"ur Astrophysik, Karl-Schwarzschild-Str.~1, 85748 Garching, Germany\label{aff136}
\and
Laboratoire Univers et Th\'eorie, Observatoire de Paris, Universit\'e PSL, Universit\'e Paris Cit\'e, CNRS, 92190 Meudon, France\label{aff137}
\and
Departamento de F{\'\i}sica Fundamental. Universidad de Salamanca. Plaza de la Merced s/n. 37008 Salamanca, Spain\label{aff138}
\and
Universit\'e de Strasbourg, CNRS, Observatoire astronomique de Strasbourg, UMR 7550, 67000 Strasbourg, France\label{aff139}
\and
Center for Data-Driven Discovery, Kavli IPMU (WPI), UTIAS, The University of Tokyo, Kashiwa, Chiba 277-8583, Japan\label{aff140}
\and
Ludwig-Maximilians-University, Schellingstrasse 4, 80799 Munich, Germany\label{aff141}
\and
Max-Planck-Institut f\"ur Physik, Boltzmannstr. 8, 85748 Garching, Germany\label{aff142}
\and
Jodrell Bank Centre for Astrophysics, Department of Physics and Astronomy, University of Manchester, Oxford Road, Manchester M13 9PL, UK\label{aff143}
\and
California Institute of Technology, 1200 E California Blvd, Pasadena, CA 91125, USA\label{aff144}
\and
Department of Physics \& Astronomy, University of California Irvine, Irvine CA 92697, USA\label{aff145}
\and
Departamento F\'isica Aplicada, Universidad Polit\'ecnica de Cartagena, Campus Muralla del Mar, 30202 Cartagena, Murcia, Spain\label{aff146}
\and
Instituto de F\'isica de Cantabria, Edificio Juan Jord\'a, Avenida de los Castros, 39005 Santander, Spain\label{aff147}
\and
Observatorio Nacional, Rua General Jose Cristino, 77-Bairro Imperial de Sao Cristovao, Rio de Janeiro, 20921-400, Brazil\label{aff148}
\and
INFN, Sezione di Lecce, Via per Arnesano, CP-193, 73100, Lecce, Italy\label{aff149}
\and
Department of Mathematics and Physics E. De Giorgi, University of Salento, Via per Arnesano, CP-I93, 73100, Lecce, Italy\label{aff150}
\and
INAF-Sezione di Lecce, c/o Dipartimento Matematica e Fisica, Via per Arnesano, 73100, Lecce, Italy\label{aff151}
\and
Institute of Cosmology and Gravitation, University of Portsmouth, Portsmouth PO1 3FX, UK\label{aff152}
\and
Department of Astronomy, University of Florida, Bryant Space Science Center, Gainesville, FL 32611, USA\label{aff153}
\and
Department of Computer Science, Aalto University, PO Box 15400, Espoo, FI-00 076, Finland\label{aff154}
\and
Instituto de Astrof\'\i sica de Canarias, c/ Via Lactea s/n, La Laguna 38200, Spain. Departamento de Astrof\'\i sica de la Universidad de La Laguna, Avda. Francisco Sanchez, La Laguna, 38200, Spain\label{aff155}
\and
Caltech/IPAC, 1200 E. California Blvd., Pasadena, CA 91125, USA\label{aff156}
\and
Ruhr University Bochum, Faculty of Physics and Astronomy, Astronomical Institute (AIRUB), German Centre for Cosmological Lensing (GCCL), 44780 Bochum, Germany\label{aff157}
\and
Department of Physics and Astronomy, Vesilinnantie 5, University of Turku, 20014 Turku, Finland\label{aff158}
\and
Serco for European Space Agency (ESA), Camino bajo del Castillo, s/n, Urbanizacion Villafranca del Castillo, Villanueva de la Ca\~nada, 28692 Madrid, Spain\label{aff159}
\and
ARC Centre of Excellence for Dark Matter Particle Physics, Melbourne, Australia\label{aff160}
\and
Centre for Astrophysics \& Supercomputing, Swinburne University of Technology,  Hawthorn, Victoria 3122, Australia\label{aff161}
\and
Department of Physics and Astronomy, University of the Western Cape, Bellville, Cape Town, 7535, South Africa\label{aff162}
\and
DAMTP, Centre for Mathematical Sciences, Wilberforce Road, Cambridge CB3 0WA, UK\label{aff163}
\and
Kavli Institute for Cosmology Cambridge, Madingley Road, Cambridge, CB3 0HA, UK\label{aff164}
\and
Department of Astrophysics, University of Zurich, Winterthurerstrasse 190, 8057 Zurich, Switzerland\label{aff165}
\and
Department of Physics, Centre for Extragalactic Astronomy, Durham University, South Road, Durham, DH1 3LE, UK\label{aff166}
\and
IRFU, CEA, Universit\'e Paris-Saclay 91191 Gif-sur-Yvette Cedex, France\label{aff167}
\and
Oskar Klein Centre for Cosmoparticle Physics, Department of Physics, Stockholm University, Stockholm, SE-106 91, Sweden\label{aff168}
\and
Astrophysics Group, Blackett Laboratory, Imperial College London, London SW7 2AZ, UK\label{aff169}
\and
Univ. Grenoble Alpes, CNRS, Grenoble INP, LPSC-IN2P3, 53, Avenue des Martyrs, 38000, Grenoble, France\label{aff170}
\and
INAF-Osservatorio Astrofisico di Arcetri, Largo E. Fermi 5, 50125, Firenze, Italy\label{aff171}
\and
Dipartimento di Fisica, Sapienza Universit\`a di Roma, Piazzale Aldo Moro 2, 00185 Roma, Italy\label{aff172}
\and
Centro de Astrof\'{\i}sica da Universidade do Porto, Rua das Estrelas, 4150-762 Porto, Portugal\label{aff173}
\and
HE Space for European Space Agency (ESA), Camino bajo del Castillo, s/n, Urbanizacion Villafranca del Castillo, Villanueva de la Ca\~nada, 28692 Madrid, Spain\label{aff174}
\and
INAF - Osservatorio Astronomico d'Abruzzo, Via Maggini, 64100, Teramo, Italy\label{aff175}
\and
Theoretical astrophysics, Department of Physics and Astronomy, Uppsala University, Box 516, 751 37 Uppsala, Sweden\label{aff176}
\and
Minnesota Institute for Astrophysics, University of Minnesota, 116 Church St SE, Minneapolis, MN 55455, USA\label{aff177}
\and
Mathematical Institute, University of Leiden, Einsteinweg 55, 2333 CA Leiden, The Netherlands\label{aff178}
\and
Institute of Astronomy, University of Cambridge, Madingley Road, Cambridge CB3 0HA, UK\label{aff179}
\and
Department of Physics and Astronomy, University of California, Davis, CA 95616, USA\label{aff180}
\and
Department of Astrophysical Sciences, Peyton Hall, Princeton University, Princeton, NJ 08544, USA\label{aff181}
\and
Space physics and astronomy research unit, University of Oulu, Pentti Kaiteran katu 1, FI-90014 Oulu, Finland\label{aff182}
\and
Center for Computational Astrophysics, Flatiron Institute, 162 5th Avenue, 10010, New York, NY, USA\label{aff183}}

 \date{\today}

 \abstract{{The well-known bimodality between star-forming discs and quiescent spheroids requires the existence of two main processes: galaxy quenching, causing the strong reduction of star formation, and morphological transformation, causing the transition from disc-dominated structures to bulge-dominated ones}. In this paper, we aim to understand the link between these two processes and their relation with the stellar mass of galaxies and their local environment. Taking advantage of the first data released by the Euclid Collaboration, covering more than $60\,\mathrm{deg}^2$ with space-based imaging and photometry, we analyse a mass-complete sample of nearly one million galaxies in the range $0.25<z<1$ with $M_\ast>10^{9.5} \, M_\odot$, using a combination of photometric and spectroscopic redshifts. We divide the sample into four sub-populations of galaxies, based on their star-formation activity (star-forming and quiescent) and morphology (disc-dominated and bulge-dominated). We then analyse the physical properties of these populations and their relative abundances in the stellar mass versus local density plane. {Together with confirming the passivity-density relation and the morphology-density relation, we find that quiescent discy galaxies are more abundant in the low-mass regime of high-density environment where $\log_{10}(1+\delta)>1.3$. At the same time, star-forming bulge-dominated galaxies are more common in field regions with $\log_{10}(1+\delta)<0.8$, preferentially at high masses.} Building on these results {and interpreting them through comparison with simulations}, we propose a scenario where the evolution of galaxies in the field significantly differs from that in higher-density environments. The morphological transformation in the majority of field galaxies takes place before the onset of quenching and is mainly driven by secular processes taking place within the main sequence, leading to the formation of star-forming bulge-dominated galaxies as intermediate-stage galaxies. Conversely, quenching of star formation precedes morphological transformation for most galaxies in higher-density environments. This causes the formation of quiescent disc-dominated galaxies before their transition into bulge-dominated ones.}

\keywords{}

   \titlerunning{Galaxy quenching and bulge formation in \Euclid}
   \authorrunning{Euclid Collaboration: F. Gentile et al.}
   
   \maketitle

\section{Introduction}
When dealing with galaxies, form and substance are tightly connected. Since the introduction of the Hubble sequence \citep{Hubble_26}, it has been well known that the morphology of galaxies strongly correlates with their physical properties (e.g. mass and star-formation rate, SFR; see e.g. \citealt{Kauffmann_03,Brinchmann_04,Wuyts_11,Huertas_24}). Throughout most of cosmic history (up to at least $z\sim3$; see e.g. \citealt{HCompany_25}), this correlation is clearly visible in the bimodal distribution of galaxies, where most of the objects belong to two main families. The first one is characterised by blue rest-frame colours, active star formation, young stellar ages, and discy morphologies. The second one, on the contrary, includes galaxies with redder colours, negligible star formation, older stellar populations, and spheroidal morphologies (see e.g. \citealt{Strateva_01,Allen_06,Q1-SP040}). The limited number of galaxies found in between these populations (see some examples of passive discy galaxies in e.g. \citealt{Masters_10} and of star-forming spheroids in e.g. \citealt{McIntosh_14}) is normally explained through the rapid evolution from one group to another. This process must include both galaxy quenching (i.e. the stopping of star formation; see e.g. \citealt{Man_18} and references therein) and morphological transformation (i.e. the transition from a disc-dominated structure to a bulge-dominated one; see e.g. \citealt{Liu_19} and references therein). However, it is not clear whether these two processes take place at the same time or not and -- in this second case -- if there is a causal connection between them.

 The common treatment of galaxy quenching relies on the distinction between two main families of processes: those related to internal factors and those depending on external ones (see e.g. \citealt{Peng_10} and references therein). The internal processes include physical mechanisms, such as feedback from massive stars or supernovae (e.g. \citealt{Ciotti_91}), as well as active galactic nuclei (AGN; e.g. \citealt{Bower_06,Croton_06}). These processes are commonly considered together in the general class of ‘mass quenching’, since they are mostly responsible for stopping of star formation in massive galaxies, regardless of their environment. The second family, instead, includes processes able to inhibit star formation by preventing the accretion of cold gas from the intergalactic medium (starvation; see e.g. \citealt{Larson_80,VanDenBosch_08}) or by depleting the gas reservoirs in galaxies (e.g. by tidal stripping or ram-pressure stripping; see e.g. \citealt{Gunn_72,Feldmann_10,Boselli_22}). All these processes are normally referred to as ‘environmental quenching’, since they are more effective in high-density environments such as galaxy groups or clusters \citep{Feldmann_10}. One of the main results of the current research in this field is that these two effects are separable (i.e. they act independently of each other) and effective at least since redshift $z\sim1$ \citep{Peng_10}. At higher redshifts, the picture becomes less clear, and a consensus in the community is still far from being reached, with several studies presenting different conclusions about the effectiveness of environmental quenching at $z>1.5$ (see, for instance, \citealt{Fossati_17,Foltz_18,Xu_25, Pan_25}).

When it comes to morphological transformation and its relationship with mass and environmental quenching, scientific literature becomes less rich.\footnote{It is interesting how taking the morphological transformation into account was already seen as a priority by \citet{Peng_10}.} While it is well established that the fraction of early-type galaxies is higher at high masses and in high-density environments (see e.g. \citealt{Dressler_80,Dressler_97,Postman_05,Q1-SP017}), the physical mechanisms able to produce the morphological transformation are still to be completely understood. A possible solution can reside in galaxy mergers, which are able to dissipate angular momentum and cause the transition from disc-dominated structures to bulge-dominated ones (see e.g. \citealt{Hopkins_08, Cox_08}), even though it is still debated if the cessation of star formation induced by mergers is permanent or not (see e.g. \citealt{Dubois_16,Athanassoula_16}). An analogous debate is still active on the possible effects of feedback mechanisms on the angular momentum of galaxies, and on the consequent possibility of inducing a morphological transformation (see e.g. \citealt{Ubler_14,Agertz_16,Yang_24}). In addition, the possibility of morphological quenching has been proposed by \citet{Martig_09}. In this scenario, the formation of a dominant bulge in star-forming galaxies could help stabilise the disc, preventing further star formation.

A full observational test of these scenarios is quite challenging, since it requires wide sky surveys (to sample different environments with different densities), with good photometric constraints (to properly assess photometric redshifts and physical properties of the observed galaxies), and high-resolution imaging (to accurately constrain the morphology of galaxies). Until now, no survey has met all these criteria. On the one hand, wide ground-based surveys missed the key ingredient of high-resolution imaging, while, on the other, space-based missions (such as the \HST or the {\it James Webb} Space Telescope) are fundamentally limited by their small fields of view, making it hard to perform large sky surveys (see some noteworthy examples in \citealt{Scoville_07} and \citealt{Casey_23}).

The recent launch of the \Euclid satellite \citep{Laureijs11,EuclidSkyOverview} promises to be a game changer in this field. Several studies based on the first data collected by the telescope already showed how -- even only relying on photometric redshifts -- these data can be employed to characterise the environment of galaxies (e.g. \citealt{Q1-SP017}), their position in the cosmic web (e.g. \citealt{Q1-SP028}) and their physical properties (e.g. \citealt{Q1-SP031,Q1-TP005}). In this study, we aim to address the scientific issues introduced before by adding morphological information to the classical studies focusing on galaxy quenching as a function of mass and environment. In our approach, we focus on the ‘intermediate stages’ in the transition between star-forming discs and quiescent spheroids: objects with either discy structures and negligible star formation or with predominant bulges and active star formation.

The paper is structured as follows. In Sect. \ref{sec:data}, we introduce the first \Euclid data used in our analysis. In Sect. \ref{sec:densty}, we describe the procedure followed to estimate the local environmental density for our galaxies. In Sect. \ref{sec:sub-populations}, we describe the four sub-populations of galaxies considered in our analysis and compute their relative abundances as a function of stellar mass and local density. The observed properties are then linked to the hosting dark matter halos through comparison with simulations. These observational results are then employed in Sect. \ref{sec:discussion} to propose an evolutionary scenario for the four sub-populations. Finally, we draw our conclusions in Sect. \ref{sec:summary} and present the future perspectives of our work in anticipation of the next data releases from the \Euclid satellite.

Throughout this paper, we assume a standard $\Lambda$CDM cosmology with the parameters reported in \citet{Planck_18}. We also assume a \citet{Chabrier_03} initial mass function. Finally, all the magnitudes quoted throughout the text are reported in the AB photometric system \citep{Oke_83}.

\section{Data}
\label{sec:data}
\subsection{The Euclid survey and the Q1 data}

During the nominal duration of its mission, the \Euclid satellite will carry out a wide survey covering almost \num{14000}$\,\mathrm{deg}^2$ of the extragalactic sky \citep{Scaramella-EP1}. As part of this survey, it will collect space-based imaging in one broad visible (VIS) filter ($\IE$; \citealt{EuclidSkyVIS}) and three near-infrared (NIR) filters ($\YE$, $\JE$, and $\HE$; \citealt{EuclidSkyNISP}), together with slitless spectroscopy at NIR wavelengths \citep{EC_Gillard}. This wide survey will be complemented by a deeper one (covering the three Euclid Deep Fields -- EDFs -- for a total of more than $50\,\mathrm{deg}^2$ at the end of the survey) with significantly higher exposure times (expected gain of about two magnitudes in photometry; see \citealt{EuclidSkyOverview}). We refer to \citet{Laureijs11} and \citet{EuclidSkyOverview} for a complete description of the \Euclid satellite, its main scientific goals, and the technical details of the telescope.

In this paper, we analyse the first data released by the Euclid Consortium (the \citealt{Q1cite}; Q1 in the following, see \citealt{Q1-TP001}). These data cover $63.1\,\mathrm{deg}^2$, divided between the deep fields Fornax (EDF-F; $12.1\,\mathrm{deg}^2$), North (EDF-N; $22.9\,\mathrm{deg}^2$), and South (EDF-S; $28.1\,\mathrm{deg}^2$). The 10 $\sigma$ depth of the Q1 data (measured in a circular aperture with a radius equal to twice the full width half maximum of the point spread function; PSF FWHM) is fixed to $\IE = 24.5$, equivalent to the nominal depth of the wide survey at the end of the mission lifetime. A complete description of the Q1 data can be found in \citet{Q1-TP001}.

For our analysis, we employ the photometric redshifts and physical properties (stellar masses, SFR\footnote{{Throughout this study, we use a SFR averaged on the last 100 Myr of the star formation history of each galaxy (see \citealt{Q1-TP005,Q1-SP031}).}}, and rest-frame colours) estimated in \citet{Q1-SP031} through the machine-learning algorithm Nearest-Neighbour Photometric Redshift (\texttt{nnpz}; see \citealt{Desprez-EP10, EP-Enia}). These data differ from the standard data products included in the Q1 release (see their descriptions in \citealt{Q1-TP002,Q1-TP003,Q1-TP004,Q1-TP005}). We refer to \citet{Q1-SP031} for a complete discussion on the main differences between the two data-sets. Here, we highlight the fact that the physical properties obtained by \citet{Q1-SP031} are computed by taking into account the \Euclid photometry ($\IE$, $\YE$, $\JE$, and $\HE$ bands), the external ground-based photometry collected between $0.3\,\micron$ and $1.8\,\micron$ by the Ultraviolet Near-Infrared Optical Northern Survey (UNIONS; \citealt{Gwyn_25}) and the Dark Energy Survey (DES; \citealt{Flaugher_15,DES_2016}), and the mid-infrared (MIR) photometry at $3.6\,\micron$ and $4.5\,\micron$ collected by \textit{Spitzer}/IRAC and extracted by \citet{Q1-SP011}. The last component is not included in the official data products by the Euclid Collaboration, even though it is available for most of the sources from the Q1 release thanks to the Cosmic Dawn
Survey of the Euclid Deep and Auxiliary Fields \citep[DAWN;][]{Moneti-EP17,EP-McPartland}. The inclusion of MIR photometry produces a significant improvement in the accuracy of the photometric redshifts and stellar masses, especially at higher redshifts (see the discussion in \citealt{Q1-SP031}).

The second tier of data products employed in this analysis consists of the morphological parameters estimated by the MERge Processing Function (MER PF; \citealt{Q1-TP004}). {Specifically, we make use of the modelling with a single bi-dimensional Sérsic profile \citep{Sersic_63} performed on the VIS images. The modelling is performed with the \texttt{SourceXtractor++} code \citep{Bertin_20,Kummel_22}, as described in more detail in \citet{Q1-SP040}.}

\subsection{Sample selection}
\label{sec:sample_selection}
As already done in several studies analysing the first \Euclid data, we apply a series of criteria to clean the sample from spurious detections. These are based on the quality flags introduced by \citet{Q1-TP005} and correspond to those employed in \citet{Q1-SP031}:

\begin{equation}
\begin{cases}
    \texttt{SPURIOUS\_FLAG}=0\ ; \\
    \texttt{DET\_QUALITY\_FLAG}<4 \ . \\
\end{cases}
\end{equation}

{These criteria are useful to avoid spurious detections and sources that are saturated or too close to the borders of the detection image}. In addition to these criteria, we aim to remove as many non-galaxies {and point sources} as possible from our sample, so we add some extra conditions following \citet{Q1-SP027} and \citet{Q1-SP040}:

\begin{equation}
\label{eq:cuts}
\begin{cases}
    \texttt{MUMAX\_MINUS\_MAG}>-2.6\ ;\\
    \texttt{PROB\_QSO}<0.86\ ; \\
    \texttt{PROB\_STAR}<0.10\ ; \\
    q>0.05\ ; \\
    0.01a<R_{\rm e}<2a\ . \\   
\end{cases}
\end{equation}

\noindent Here, $q$ and $R_{\rm e}$ are the axis ratio and effective radius of the Sérsic profile, $a$ is the major semi-axis of the segmentation area, {\texttt{MUMAX\_MINUS\_MAG} is a compactness criterion defined as the difference between the peak surface brightness and the total magnitude in the detection band, and \texttt{PROB\_QSO} and \texttt{PROB\_STAR} quantify the probability of a given object to be a quasi-stellar object (QSO) or a star, respectively. As reported in \citet{Q1-TP005}, the QSO classification is obtained through a random forest classifier and achieved a 95\% success rate when validated on a spectroscopic sample of non-stellar objects. This value increases slightly when the threshold adopted in Equation \ref{eq:cuts} is chosen, as reported by \citet{Q1-SP027}.}

Finally, we aim at analysing a complete sample of galaxies in the redshift range $0.25<z<1$. Therefore, we choose to work with a flux- and mass-limited sample. For doing so, we perform a last series of cuts:

\begin{equation}
\label{eq:criteria}
\begin{cases}
    \HE<24\ ; \\
    M_\ast>10^{9.5} \, {M_\odot};\\
    0.25<z_{\rm phot}<1\ .
\end{cases}
\end{equation}

{The first condition roughly corresponds to a S/N>5 cut in the $\HE$ band, which allows us to focus on a sample of galaxies with good constraints in the NIR regime}. The chosen limiting mass corresponds to the $95\%$ completeness limit for quiescent galaxies with $\HE<24$ at $z\sim1$, as estimated by \citet{Q1-SP031} following the standard relation of \citet{Pozzetti_10} applied to their data products. {Given the overall behaviour of the mass completeness, the percentage is clearly higher at lower redshifts and for star-forming galaxies.} The chosen redshift range allows us to focus on galaxies covering a significant cosmological volume and with an accurate photometric redshift (see Sect. \ref{sec:densty}). The final sample of galaxies employed to estimate the density field (Sect. \ref{sec:densty}) includes \num{975830} objects. The rest of the analysis, where we intensively take advantage of the morphological parameters, is based on a sub-sample of \num{876727} galaxies with
\begin{equation}
\begin{cases}
    q<1\ ;\\
    0.302<n_{\rm Ser}<5.45\ ,
\end{cases}
\end{equation}
where $n_{\rm Ser}$ is the Sérsic index. These cuts are prescribed by \citet{Q1-SP040} to remove from the sample galaxies with inaccurate morphological analysis.

\begin{figure*}
    \centering
    \includegraphics[width=0.243\linewidth]{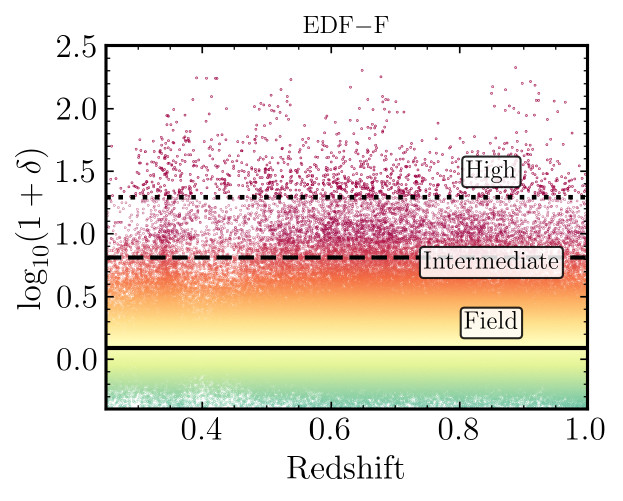} \ \includegraphics[width=0.243\linewidth]{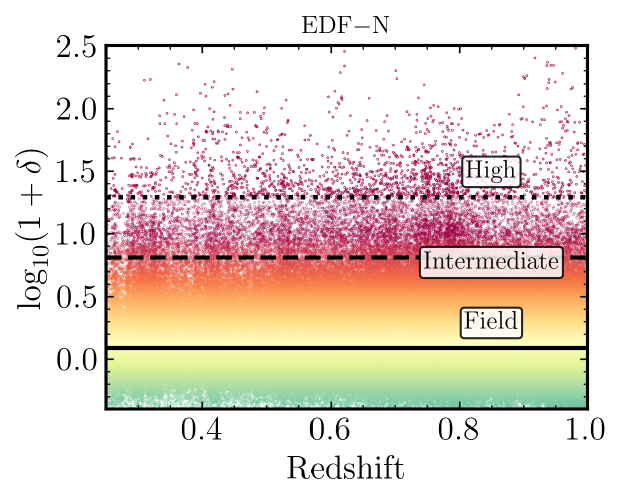} \ \includegraphics[width=0.243\linewidth]{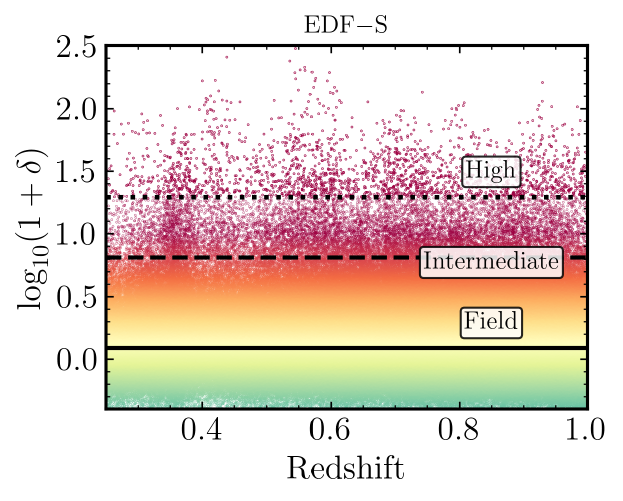}   \ \includegraphics[width=0.243\linewidth]{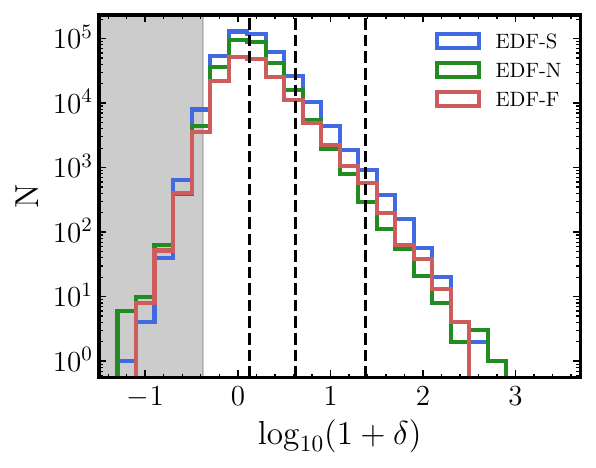}  
    \caption{\textit{(First three panels)} Density field as estimated in Sect.\,\ref{sec:densty} for the three EDFs. The three panels show the density contrast parameter as a function of the redshift in each field. The horizontal solid, dashed, and dotted lines represent the median value of $\log_{10}(1+\delta)$, the {\rev $2\sigma$, and the $5\sigma$ levels, respectively}. They separate the field, intermediate-density regions, and high-density ones. \textit{(Last panel)} Distribution of the $\log_{10}(1+\delta)$ in the three EDFs. The dashed lines indicate the same levels as in the previous panels. The shaded area reports the area excluded from the analysis.}
    \label{fig:density_field}
\end{figure*}

\subsection{Ancillary spectroscopy}
\label{sec:specz}

To incorporate the spectroscopic information available for some of the galaxies observed in Q1, we take advantage of the spectroscopic compilation available within the Euclid Consortium (C. Saulder, private communication). This collection includes redshifts from: the Dark Energy Spectroscopic Instrument \citep[DESI,][]{2016arXiv161100036D, 2024AJ....168...58D}; the 16th Data Release of the Sloan Digital Sky Survey \citep[SDSS,][]{2020ApJS..249....3A}; the 2MASS Redshift Survey \citep[2MRS,][]{2012ApJS..199...26H}; the PRIsm MUlti-object Survey \citep[PRIMUS,][]{2011ApJ...741....8C}; the Australian Dark Energy Survey \citep[OzDES,][]{2015MNRAS.452.3047Y, 2017MNRAS.472..273C, 2020MNRAS.496...19L}; 3dHST \citep{2012ApJ...758L..17B}; the 2-degree Field Galaxy Redshift Survey \citep[2dFGRS,][]{2001MNRAS.328.1039C}; the 6-degree Field Galaxy Redshift Survey \citep[6dFGS,][]{2009MNRAS.399..683J}; the MOSFIRE Deep Evolution Field Survey \citep[MOSDEF,][]{2015ApJS..218...15K}; the VANDELS ESO public spectroscopic survey \citep{2018arXiv181105298P, 2023A&A...678A..25T}; the JWST Advanced Deep Extragalactic Survey DR3 \citep[JADES,][]{2025ApJS..277....4D}; the 2-degree Field Lensing Survey \citep[2dFLens,][]{2016MNRAS.462.4240B}; and the VIMOS VLT deep survey \citep[VVDS,][]{2005A&A...439..845L}. These are all matched to \Euclid sources. We only include in our procedure sources with secure redshifts (i.e. with quality flag $Q_{\rm f}\ge3)$, obtaining a spectroscopic coverage of about $5\%$ of the galaxies involved in the analysis, decreasing quite linearly from $10\%$ at $z=0.25$ to $2\%$ at $z=1$. {In the considered redshift range, we find an overall good agreement between the photometric and the spectroscopic redshift, with a $\sigma_{\rm \Delta z/(1+z)}\sim 0.3$. {\rev We underline, however, that the spectroscopic coverage is biased towards star-forming galaxies, for which the presence of emission lines makes it easier to model the spectroscopic redshift.}

\section{Estimation of the density field}
\label{sec:densty}

We reconstruct the density field of the Q1 galaxies following a tomographic approach, necessary because the accuracy of the available photometric redshifts does not allow a 3D modelling ($\sigma_{\Delta z/(1+z)}\sim0.03$; see e.g. the discussion in \citealt{Malavasi_16} on the possibility of reconstructing the density field with such photometric redshifts). Following \citet{Q1-SP028}, we divide the redshift range $0.25<z<1$ into 20 overlapping redshift slices with a fixed comoving width of $170\,\mathrm{Mpc}\,h^{-1}$. The overlap is constant and fixed to $90\,\mathrm{Mpc}\,h^{-1}$. The choice of a fixed comoving size for the redshift slices will allow us to compare our results in different redshift bins. Moreover, by employing overlapping slices, we reduce the probability of missing significant over-densities because of the slicing. The lower bound of the redshift range is chosen to sample a large enough volume {(and, hence, a large enough sample size)}, while the upper bound is justified by the decreasing accuracy of the photometric redshifts and by the necessity of analysing a mass-complete sample. Moreover, in this redshift range the VIS filter samples a narrow range of rest-frame optical emission, avoiding biases in the determination of the morphologies (see Sect. \ref{sec:sub-populations}).

In each redshift slice, we reconstruct the density field through the $\Sigma_{\rm N}$ estimator (see e.g. \citealt{Postman_05,Baldry_06,Q1-SP017}), defined as
\begin{equation}
    \Sigma_N=\frac{N+1}{\pi R_N^2}\ ,
\end{equation}
where $R_ N$ is the projected distance from the considered galaxy to the $N$-th closest neighbour. Since the density field is mostly traced by massive galaxies (which, being brighter, are also characterised by a more robust photo-\textit{z}; $\sigma_{\Delta z/(1+z)}=0.015$), we compute the $\Sigma_N$ estimator for the galaxies with $M_\ast>10^{10.3} \, M_\odot$ (see e.g. \citealt{Q1-SP028}) and then assign to the less massive galaxies the density field value of the closest massive neighbour. {We underline that our sample is mass-complete at all redshifts with this choice (see Sect. \ref{sec:sample_selection}}).

To properly take into account the uncertainties on the photometric redshifts, we rely on a Monte Carlo integration. In practice, we perform $10^3$ realisations of the density map of each redshift slice in each of the three EDFs covered by Q1. In each iteration, we randomly sample a value from the Gaussianised probability distribution of the photo-\textit{z}s for each galaxy. For the galaxies for which a spectroscopic redshift is available (see Sect.\,\ref{sec:specz}), we model the probability distribution as a delta function centred on the spec-\textit{z} value.

This procedure allows us to estimate the posterior probability distribution of $\Sigma_N$ for all the galaxies in our sample. In the following, we will assume as the reference value of $\Sigma_N$ the median value of the posterior distribution and as the related uncertainty half the symmetrised interval between its 84th and 16th percentiles. Moreover, accounting for the evolution of the mean density of the Universe, in the rest of the analysis we will employ the density contrast parameter, defined as
\begin{equation}
    \log_{10}(1+\delta)=\log_{10}\left(1+\frac{\Sigma_N - \bar{\Sigma}_N}{\bar{\Sigma}_N}\right)\ ,
\end{equation}
where $\bar{\Sigma}_N$ is the median density of the galaxy sample in a given redshift slice. To avoid spurious effects due to galaxies close to the edges of the deep fields and of masked regions, we do not take into account any object with a density contrast parameter lower than $\bar\Sigma_N-2\sigma$, with $\sigma$ being the standard deviation of the distribution estimated through sigma-clipping.

Finally, we perform the procedure discussed above for several values of $N$, in the range $[3$,$10]$. Since we notice that the widest dynamical range of $\log_{10}(1+\delta)$ is achieved with $N=5$, in the rest of the analysis we will only focus on $\Sigma_5$. An example of the density field as a function of redshift in the three EDFs is reported in Fig.\,\ref{fig:density_field}. The estimated density field shows a logarithmically Gaussian shape, with median and standard deviation not evolving strongly with redshift {\rev or with the position within the EDFs}. {\rev Therefore, we highlight three different regimes: in the following we will refer to ‘field galaxies’ as all the sources with a $\log_{10}(1+\delta)$ below ${\rm median}+2\sigma$, galaxies in ‘intermediate density regions’ where ${\rm median}+2\sigma<\log_{10}(1+\delta)<{\rm median}+5\sigma$, and ‘high density’ where $\log_{10}(1+\delta)>{\rm median}+5\sigma$. A quantitative assessment of the reliability of these definitions can be found in the Appendix \ref{app:completeness}.}

\begin{figure*}
    \centering
    \includegraphics[width=0.49\linewidth]{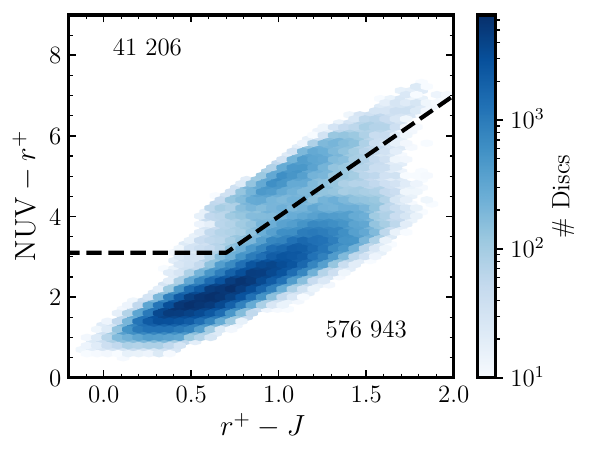} \, \includegraphics[width=0.49\linewidth]{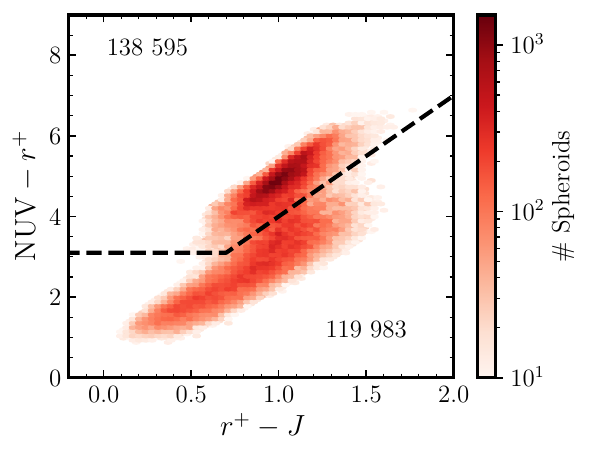}   
    \caption{Selection of star-forming and quiescent galaxies following the rest-frame $\mathrm{NUV}$--$r^{+}$--$J$ colour selection by \citet{Ilbert_10}, with the quiescent galaxies located in the upper left part of the plot. The left panel reports the number of discy galaxies belonging to the two selected families, the right panel the number of spheroidal galaxies. The numbers in the two plots report how many galaxies belong to each of the four sub-populations.}
    \label{fig:classification}
\end{figure*}

\section{Sub-populations of galaxies}
\label{sec:sub-populations}
\subsection{Classification scheme}

Our goal is to analyse the interplay between star formation, stellar mass, environment, morphology, and redshift. Clearly, even with the large statistics offered by the Q1 data, it is extremely challenging to analyse the correlations between so many continuous variables in a multi-dimensional space. For this reason, we simplify the problem by studying the evolution of different sub-populations of galaxies as a function of stellar mass and environment in different redshift bins. 

The starting point is the bi-modality reported by \citet{Q1-SP040} in the rest-frame $u-r$ versus Sérsic index plane, which divides the galaxies into red and blue, and into disc-dominated and bulge-dominated.\footnote{For the sake of brevity, in the following we will refer to these populations as ‘discs’ and ‘spheroids’, acknowledging the fact that both components are present in all the objects, with different ratios.} Instead of dividing the sample into the two classical families of early- and late-type galaxies (where the colour information is combined with the morphology), we rely on a classification in four families of objects. The first classification regarding the morphology of galaxies is the same as in \citet{Q1-SP040}, employing $n_{\rm Ser}=2$ as the threshold between disc- and bulge-dominated galaxies. 

For the second classification into star-forming and quiescent galaxies, we employ a combination of two rest-frame colours ($\mathrm{NUV}-r^{+}$ and $r^{+}-J$) to have a more robust criterion against contamination by dusty star-forming galaxies than by only using a single $u-r$ colour. We underline that this precaution is needed since the dust content of galaxies is expected to correlate with mass (see e.g. \citealt{Salim_20} and references therein), which is one of the variables involved in our analysis. We adopt the selection by \citet{Ilbert_10}, classifying as quiescent all the sources with both

\begin{equation}
\label{eq:quiescent}
\begin{cases}
    \mathrm{NUV} - r^+>3(r^+-J) +1\ ,\\
    \mathrm{NUV} - r^+>3.1\ .
\end{cases}
\end{equation}

We note that this selection is the same employed by \citet{Q1-SP031} to compute the mass completeness of the sample; this ensures that our analysis is performed on a $95\%$ mass-complete sample. A visual representation of the selection in Eq. (\ref{eq:quiescent}) is reported in Fig.\,\ref{fig:classification}.

The above-mentioned classifications allow us to separate the full sample of galaxies into four populations. Two of them represent the more common stages of galaxies in the Universe at $z<1$: the star-forming discs and the quiescent spheroids. These two classes encompass around $82\%$ of the total sample of galaxies at $0.25<z<1$ (with $66\%$ of star-forming discs and $16\%$ of quiescent spheroids). The other two classes are normally considered as intermediate phases of the evolution between the two main categories: the quiescent discs and the star-forming spheroids. This intermediate scenario is suggested by their low number densities ($\sim18\%$ of the total, with 5\% of quiescent discs and 13\% of star-forming spheroids). Some examples of galaxies belonging to the four families are shown in Fig.\,\ref{fig:examples}. We underline that the choice of a single threshold for dividing the two populations is clearly an over-simplification: we expect the population of spheroids to include both objects where the stellar disc is completely absent, as well as others where its contribution is less significant than that of the bulge (e.g. Sa spirals and lenticular galaxies). {The proposed classification is affected by two main sources of uncertainty. On the one hand, those coming from the estimation of the physical parameters (i.e. the rest-frame colours) and -- consequently -- the classification in star-forming and quiescent galaxies. On the other hand, the uncertainty on the Sérsic modelling and the consequent classification in discy and spheroidal galaxies. We explore the effect of these uncertainties on our classification scheme through a Monte Carlo integration, where the four classes are recomputed $10^3$ times. For each iteration, a value for the three rest-frame colours and for the Sérsic index is randomly sampled from the Gaussianised posterior probability contained in the catalogues. We report variations of less than 1\% in the different classes, ensuring that our classification scheme is robust against the considered uncertainties.}

\begin{figure*}
    \centering
    \includegraphics[width=\linewidth,trim={0 3.6cm 0 0},clip]{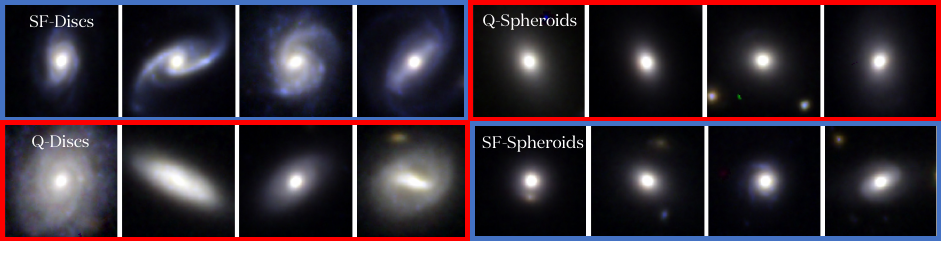}
    \caption{Some examples of galaxies selected according to the criteria presented in Sect.\,\ref{sec:sub-populations}. Starting from the upper left row and proceeding clockwise, the figure shows star-forming discs, quiescent spheroids, star-forming spheroids, and quiescent discs. All the cutouts have a 5 arcsec side and are realised by combining the \Euclid images in the \HE, \YE, and \IE filters through the algorithm by \citet{Lupton_04}.}
    \label{fig:examples}
\end{figure*}

\begin{figure*}
    \centering
    \includegraphics[width=\linewidth]{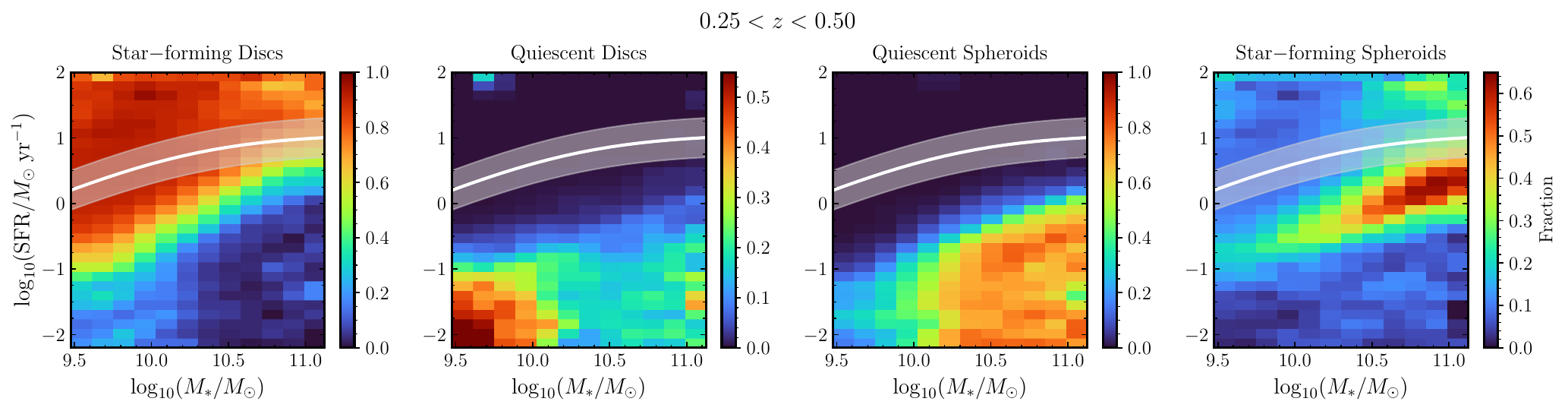} \, \includegraphics[width=\linewidth]{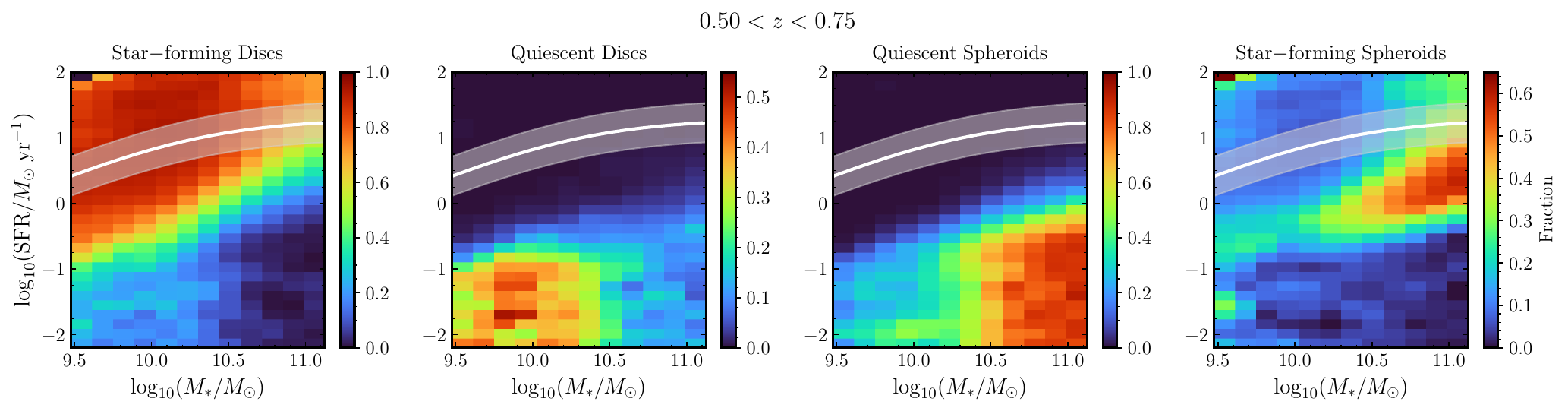} \, \includegraphics[width=\linewidth]{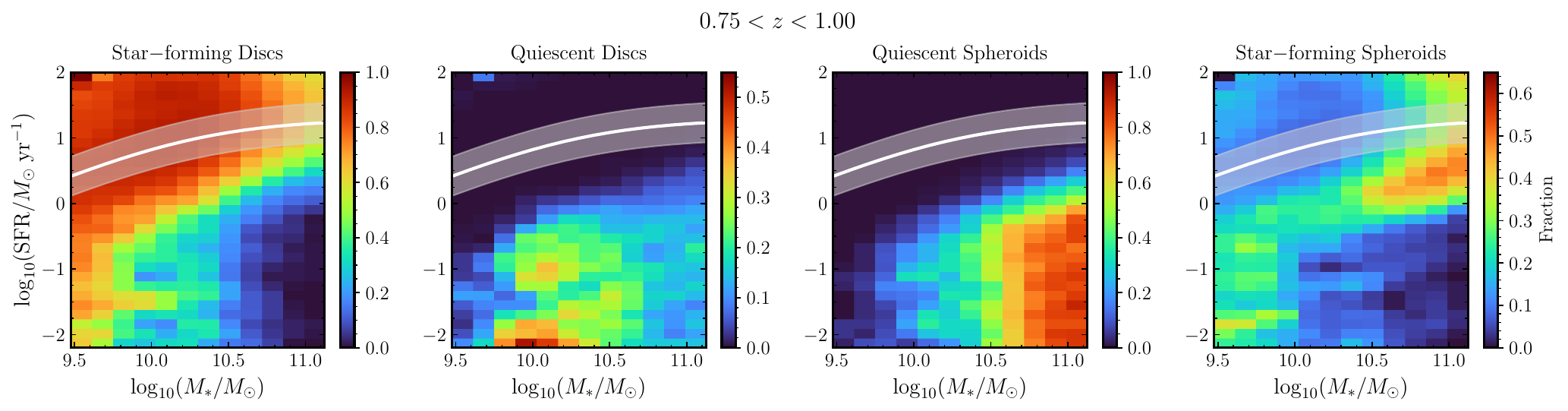}   
    \caption{Four sub-populations of galaxies in the SFR versus stellar mass plane. {Each pixel reports the relative abundance of each population through the colour-code reported in the colour-bar (which is different for each panel).} The white solid line reports the location of the main sequence of star-forming galaxies as parametrised by \citet{Q1-SP031}, while the shaded area reports its intrinsic scatter of $\sigma\sim0.3\,\mathrm{dex}$.}
    \label{fig:main_sequence}
\end{figure*}

\subsection{Population-level properties}
\label{sec:MS}

A first characterisation of the four sub-populations of galaxies relies on their physical properties. In Fig.\,\ref{fig:main_sequence}, we report the relative abundances of the four classes in the stellar mass versus star-formation rate plane in three redshift bins: $[0.25$, $0.5], [0.5$, $0.75],$ and $[0.75$, $1.0]$. Each pixel reports the relative abundance of one population in the combined bin of redshift, stellar mass, and SFR. The location of the main sequence of star-forming galaxies (MS; see e.g. \citealt{Elbaz_11,Schreiber_15,Popesso_23}) as parametrised by \citet{Q1-SP031} by analysing the same data included in this paper is shown for reference. 

We find that star-forming discs represent the vast majority of the galaxies in the MS, while the quiescent spheroids populate the so-called red cloud, with SFR at least one dex lower than MS galaxies, and with masses above $10^{10.5}\, M_\odot$. Looking at the two intermediate populations, we see that the quiescent discs represent the majority of galaxies located below the main sequence (at the same distance from the relation by \citealt{Q1-SP031} as the quiescent spheroids) with stellar masses below $10^{10.5} \, M_\odot$\footnote{A relative majority of low-mass star-forming discs in the highest redshift bin is also visible in Figure \ref{fig:main_sequence}. However, this feature is likely a spurious effect due to the low photometric S/N of the galaxies in that region of the diagram. Therefore, we do not include this feature among the results discussed in the remainder of the paper.}. The star-forming spheroids, instead, represent the majority of galaxies located in the so-called ‘green valley’ (see e.g. \citealt{Bell_04,Schawinski_14}) between the main sequence and the red cloud, at stellar masses higher than $10^{10.5} \, M_\odot$. These results confirm the well-known correlation between morphology and location in the stellar mass versus SFR plane (e.g. \citealt{Wuyts_11,Huertas_24}), as well as the mass separation between disc-dominated and bulge-dominated galaxies in the red sequence (e.g. \citealt{Quilley_22}). Moreover, the same study by \citet{Quilley_22} found that the green valley is mostly composed of bulge-dominated objects such as Sa spirals and lenticular galaxies, in good agreement with what is reported in Fig.\,\ref{fig:main_sequence}. {We also find an interesting difference between our results and those reported by \citet{Wuyts_11}: at low masses, we report a higher fraction of quiescent discs, More in detail, we report a median Sérsic index lower than 2 (see also Fig.\,12 by \citealt{Q1-SP031}), while \citet{Wuyts_11} finds a higher value for the galaxies located below the main sequence at $M_\ast<10^{10.5} \, M_\odot$. However, this tension disappears when we only include galaxies located in the field (see the definition in Sect. \ref{sec:densty}) in our analysis. This result will be explored in more detail in the next section.}

\subsection{Effects of mass and environment}
\label{sec:analysis}
\begin{figure*}
    \centering
    \includegraphics[width=\linewidth]{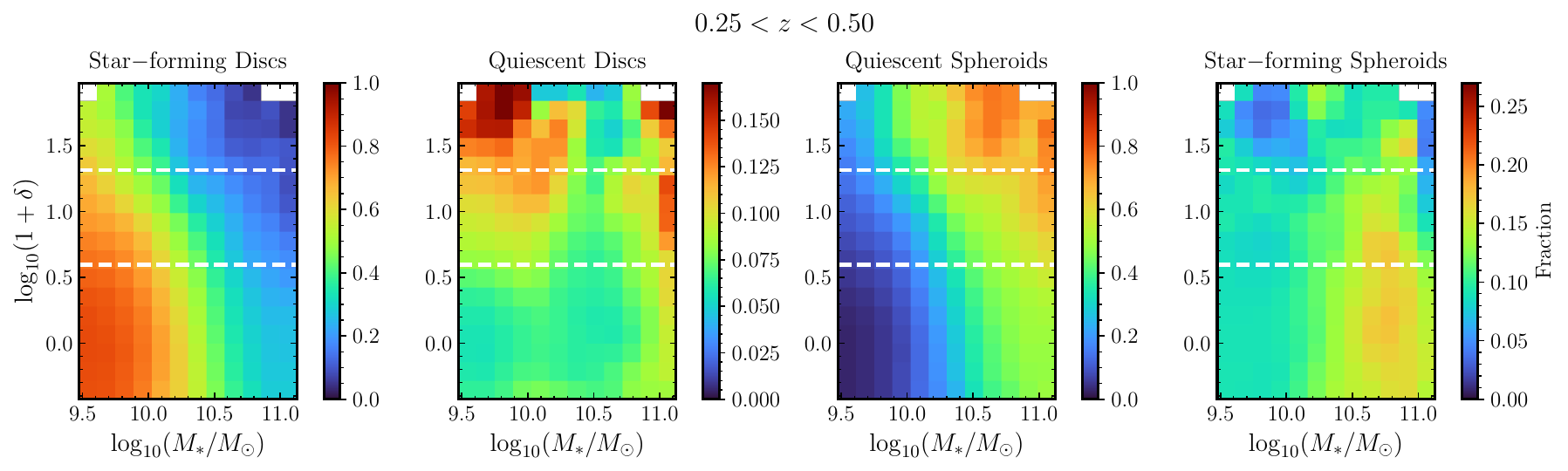} \, \includegraphics[width=\linewidth]{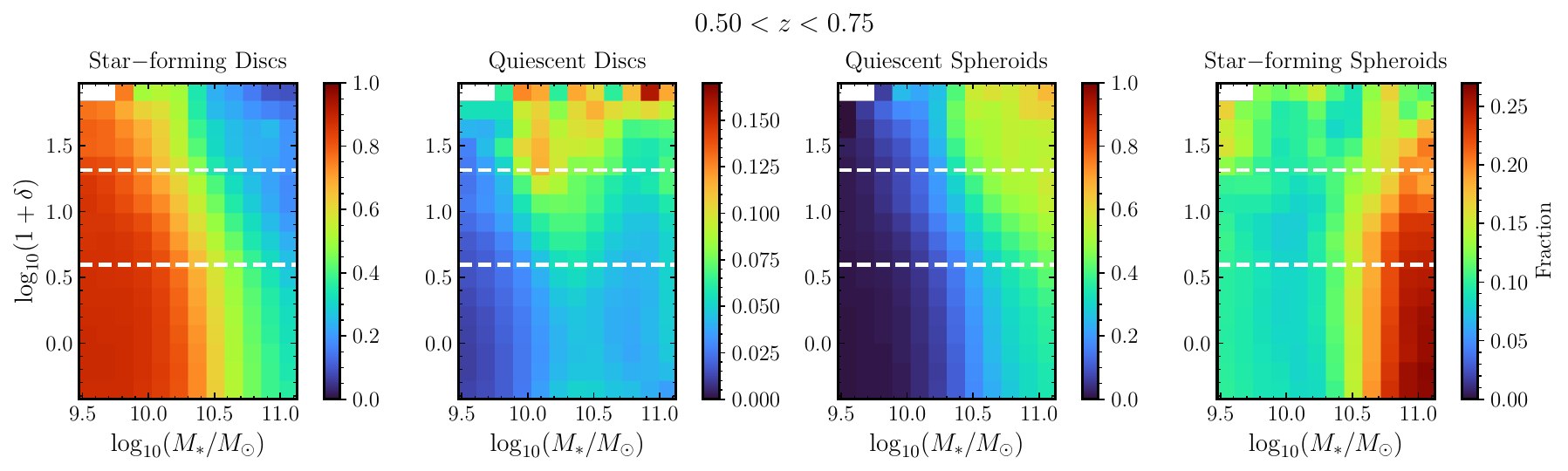} \, \includegraphics[width=\linewidth]{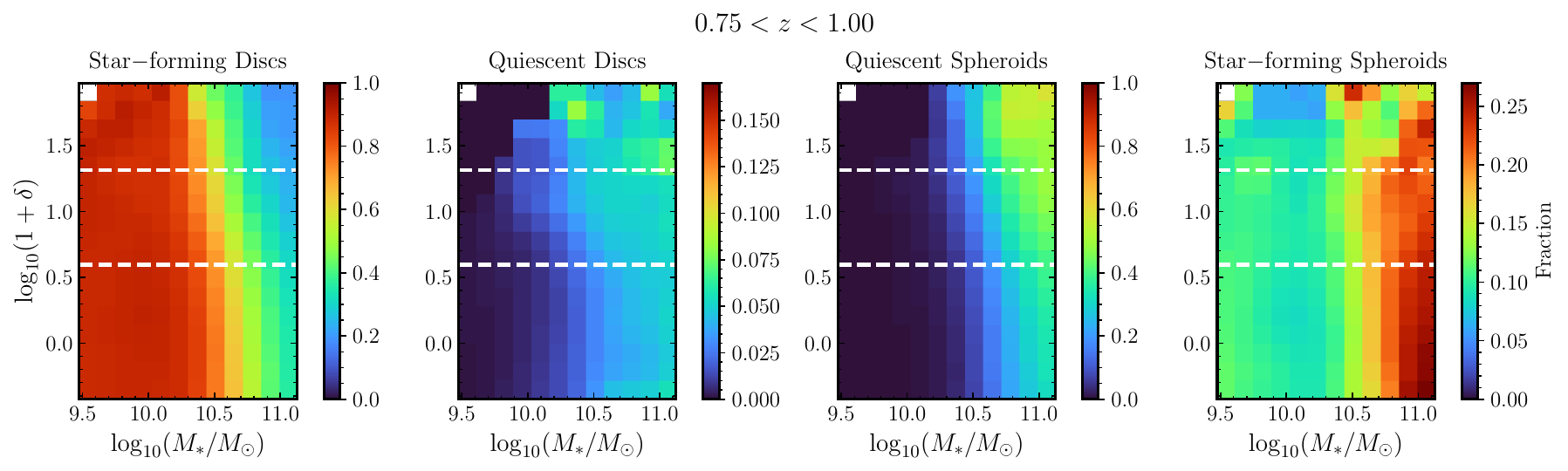}   
    \caption{Relative abundances of the four sub-populations of galaxies analysed in this paper as a function of stellar mass and local density contrast $\log_{10}(1+\delta)$. Each bin has a fixed size of $0.175\,\mathrm{dex}$ on both axes. The three rows report the results in three redshift bins with a fixed size of $0.25$ in the range $0.25<z<1$. The dashed lines separate the different density regions (field, intermediate, and high) defined in Sect. \ref{sec:densty} and Fig.\,\ref{fig:density_field}.}
    \label{fig:peng_final}
\end{figure*}

A second characterisation of the four sub-populations relies on their location in the stellar mass versus local density contrast plane, as shown in Fig.\,\ref{fig:peng_final}. These plots are realised by dividing the parameter space into several bins of mass and $\log_{10}(1+\delta)$, with a fixed logarithmic width of $0.175\,\mathrm{dex}$ on both axes. Each pixel represents the relative abundance of each population with respect to the total number of galaxies in the combined bin of redshift, stellar mass, and local density contrast. To ensure the representativeness of our results, we only report in Fig.\,\ref{fig:peng_final} the bins including at least 15 galaxies. To take into account the uncertainties on the stellar masses and on the characterisation of the environment, all the reported plots are obtained through a Monte Carlo integration, where each plot is realised $10^3$ times by sampling each time a different value from the Gaussianised posterior distribution of the stellar masses and the density contrast parameter (see Sect.\,\ref{sec:densty}). The reported plots are obtained as the median of the different realisations. We only report in the figure the bins where the signal-to-noise ratio (i.e. the median value of the bin divided by the relative uncertainty estimated as the half-symmetrised interval between the 84th and 16th percentile of the posterior distribution) is higher than three. With this choice, we limit the impact of galaxies with poorly constrained local density.

The main observational results of the plots shown in Fig.\,\ref{fig:peng_final} are the following. Firstly, we confirm the main findings of several studies such as \citet{Peng_10} and \citet{Q1-SP017}: star-forming discs represent the vast majority of galaxies at low masses and in low-density environments, while quiescent spheroids dominate the high-mass and high-density regimes. The transition point (i.e. where the star-forming discs become less than $50\%$ of the total number of galaxies) in the field (i.e. where the effect of environmental quenching is negligible) is located around $10^{10.5}\,M_\odot$, and shows a slight increase with increasing redshift. Both results are in good agreement with what was found by \citet{Peng_10} in their analysis of a spectroscopic sample of galaxies in the Sloan Digital Sky Survey and the zCOSMOS survey, ensuring the reliability of our methods. Analogously, the effects of the environment start to become visible in intermediate and high-density environments {\rev (i.e. where the density contrast is above $2\sigma$ of the density field, see the horizontal lines in Fig.\,\ref{fig:peng_final})}. The effect of environmental quenching is visible as oblique transitions instead of vertical ones at intermediate and high densities (i.e. at fixed masses below $10^{10.5}\ M_\odot$ the fraction of quiescent galaxies increases with increasing local density) and also shows an evolution with redshift, with a strength that seems to be much less significant at higher redshift. This result is also in good agreement with what is found by \citet{Peng_10} and \citet{Q1-SP017}, suggesting a lower efficiency of environmental quenching at $z>1$. A possible explanation of this result resides in the evolution of the dynamical state of over-densities with cosmic time (see e.g. \citealt{Chieng_17}).

Secondly, we observe that quiescent discs represent only a tiny fraction (less than $5\%$) of the galaxies in the field in the highest redshift bin ($0.75<z<1$), with only a slight evolution with cosmic time. Moreover, their relative abundance in the field is found to be independent on the stellar mass.\footnote{A small effect is visible in the highest redshift bin, but the overall change in relative abundance is less than $5\%$ and -- therefore -- not significant in a $95\%$ mass-complete sample, see Sect.\,\ref{sec:data}.} On the other hand, the fraction of these sources is higher in intermediate- and high-density environments, with a strong evolution with cosmic time and with a dependence on stellar mass becoming more evident and complex at lower redshifts. All these findings suggest that the formation of quiescent discs is mainly driven by environmental effects, while their possible transition into quiescent spheroids is mainly due to internal effects, as it will be discussed in detail in Sect.\,\ref{sec:GEV_Dense}.

The final observational result concerns the star-forming spheroids. These sources represent a significant fraction (up to $25\%$) of the galaxies in the high-mass end of the distribution in the field. Their relative abundance in this region of the mass-density plane evolves with cosmic time, decreasing down to less than $20\%$ in the lowest redshift bin ($0.25<z<0.5$). On the other hand, these sources are initially (i.e. at $z>0.75$) found also in over-dense regions (with fractions reaching 20--$25\%$), but then their relative abundance in intermediate- and high-density regions decreases with cosmic time, reaching levels around $10\%$. These findings suggest that the formation of star-forming spheroids is mainly driven by internal processes and uncommon in over-dense regions, as discussed in greater detail in Sect.\,\ref{sec:GEV_Field}.

\begin{figure}
    \centering
    \includegraphics[width=\columnwidth]{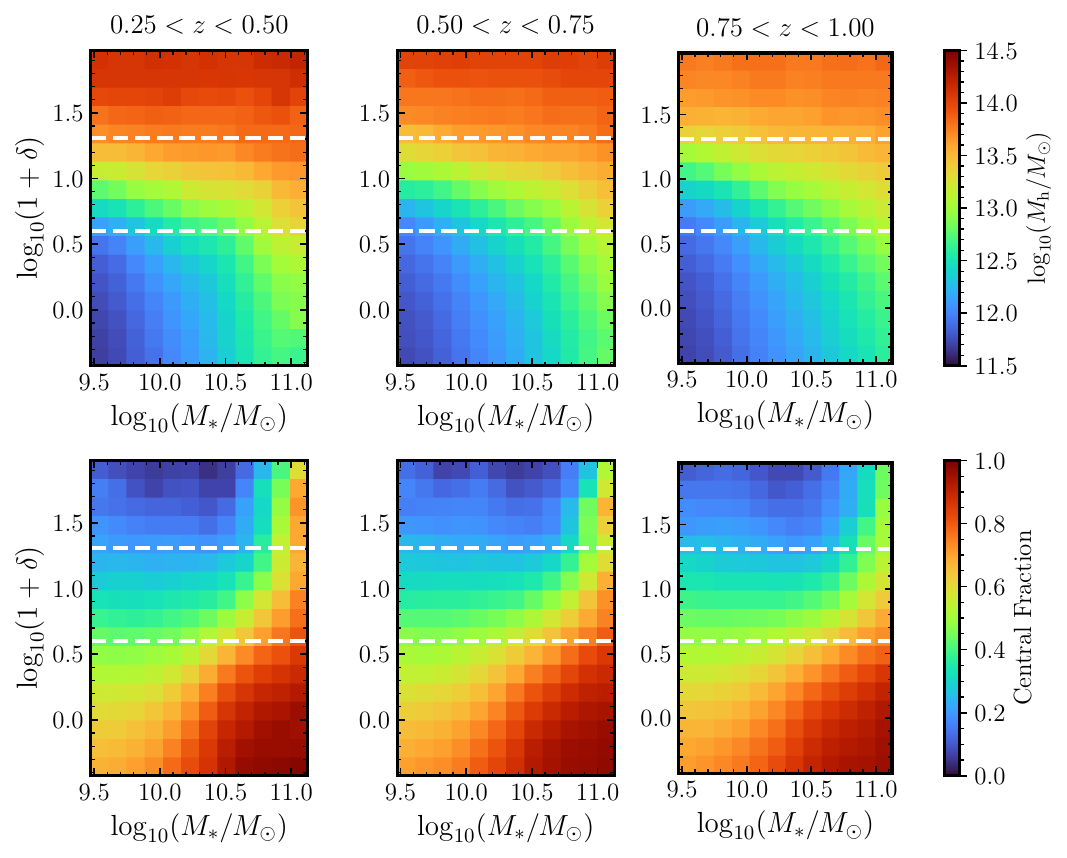}
    \caption{Median halo mass (\emph{top row}) and fraction of central galaxies (\emph{bottom row}) as a function of stellar mass and local density contrast $\log_{10}(1+\delta)$. The plot is constructed following the same procedure employed for Fig.\,\ref{fig:peng_final}, but analysing the galaxies included in the \Euclid \texttt{Flagship-2} simulation \citep{EuclidSkyFlagship}. {For satellite galaxies, the plot reports the mass of the main dark matter halo.} The horizontal dashed lines report the same density regions as in Fig.\,\ref{fig:density_field}.}
    \label{fig:simulations}
\end{figure}

 \begin{figure*}
\sidecaption
  \includegraphics[width=12cm]{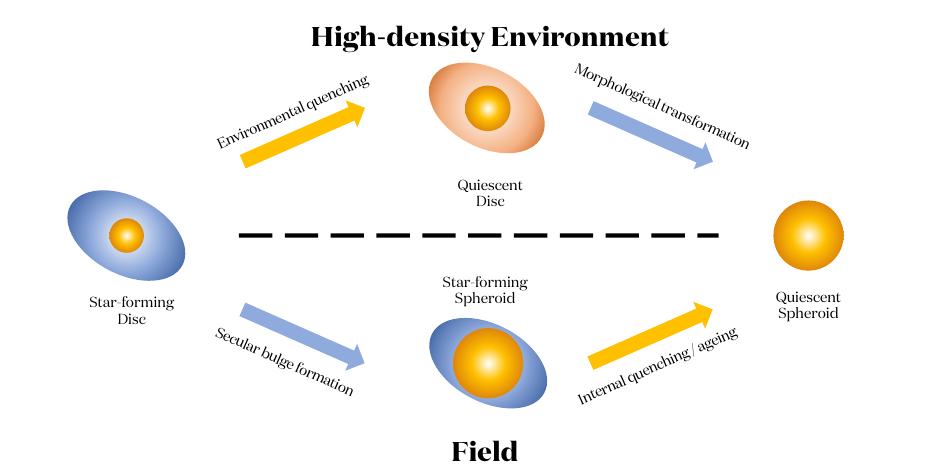}
     \caption{Sketch summarising the evolutionary scenario inferred from our observational results. The transition from star-forming discs into quiescent spheroids is significantly different in the field and in high-density environments. In the first case, the morphological transformation for most objects happens through secular evolution taking place in the main sequence and it is then followed by quenching by internal processes. In over-dense regions, on the other hand, the quenching for the majority of galaxies is due to external processes and precedes the morphological transformation. Further details in Sect.\,\ref{sec:discussion}.}
     \label{fig:model}
\end{figure*}

\subsection{Connection to dark matter halos}
\label{sec:simulations}

To offer a physical interpretation of the observational results presented in Sect. \ref{sec:analysis}, we rely on the \Euclid \texttt{Flagship-2} simulation \citep{EuclidSkyFlagship}\footnote{The \texttt{Flagship-2} simulation was retrieved from the CosmoHub portal \citep{Carretero_17,Tallada_20}: \url{https://cosmohub.pic.es/home}.}. We analyse a light-cone with the same observational properties as the sample introduced so far. In more detail, we focus on an area of $63 \ {\rm deg}^2$, selecting a mass-limited sample of galaxies following the same criteria introduced in Eq.  (\ref{eq:criteria}). For these sources, we characterise the environment using the same procedure described in Sect. \ref{sec:densty}, by employing the same redshift slicing introduced there. Since we are dealing with a simulation, we have access to additional parameters that we could not include in our analysis. Specifically, we have access to the masses of the dark matter halos hosting our sources and to the distinction between central and satellite galaxies (see \citealt{EuclidSkyFlagship}). Following the same procedure as in Sect. \ref{sec:analysis}, we obtain the plots shown in Fig.\,\ref{fig:simulations}. There, we report the median halo mass and the fraction of central galaxies as a function of the stellar mass and local density contrast, in the same three redshift bins introduced in the previous sections. Looking at the upper panels of Fig.\,\ref{fig:simulations}, it is possible to notice how the halo mass tends to increase monotonically with the stellar mass for the galaxies in the field, while it is almost independent on the stellar mass for galaxies in over-dense regions. Moreover, this uniform halo mass for the sources in these latter regions increases with cosmic time (as expected given the evolution of the halo mass function; see e.g. \citealt{Watson_13}). Looking at the bottom panels of the same figure, we can see how the field is dominated by central galaxies, while the over-dense regions present a clear bi-modality. Specifically, galaxies less massive than $M_\ast\sim10^{10.5}\ M_\odot$ are mainly satellites, while the higher masses are dominated by central galaxies. {We will use these findings in the next section to develop a physical interpretation of our observational results.}

\section{Evolutionary implications}
\label{sec:discussion}

In this section, we focus on the possible evolutionary implications of the observational results presented in the previous sections. Based on this evidence, we propose a simple evolutionary scenario where galaxies start their evolution as discs, they evolve as part of the main sequence, then they quench their star formation and reach the red cloud {(see also e.g. the analogous studies by \citealt{Bouche_10}, \citealt{Wuyts_11}, or \citealt{Davies_25})}. This simple scenario does not include any additional phenomena such as rejuvenation (e.g. \citealt{Chauke_19,Mancini_19,MartinNavarro_22}) or wet major mergers. The first phenomenon is hard to take into account without proper stellar ages for our objects, but -- anyway -- it is not expected to impact more than $\sim20\%$ of galaxies in the mass and redshift ranges covered by our observations (see e.g. \citealt{Chauke_19,ArangoToro_25}, but also a different perspective by \citealt{Mancini_19}). Similarly, the impact of wet major mergers on the morphology is expected to be significant only at very high masses ($M_\ast>10^{11} M_\odot$; see e.g. \citealt{RodriguezGomez_17}). At the same time, our scenario aims to explain our observations at $z<1$; therefore it does not account for the possible direct formation of star-forming compact galaxies through -- for instance -- feedback-free mechanisms (\citealt{Dekel_23}) expected to take place at higher redshifts.

In our scenario, we will assume that the effects of the environment are negligible in the field (in our analysis, where the density contrast parameter is below $+2\sigma$ from its median value). We acknowledge that this last assumption represents an oversimplification and that a more complete treatment of the problem would require the assembly of a sample of galaxies in cosmic voids (see e.g. \citealt{Kreckel_12} and Euclid Collaboration: Papini et al., in prep.).

In this section, we will focus on the formation paths of the two intermediate populations, namely the star-forming spheroids and the quiescent discs as proxies to investigate galaxy quenching and morphological transformation in environments with different densities. A sketch summarising our scenario is shown in Fig.\,\ref{fig:model}.

\subsection{Galaxy evolution in the field and role of star-forming spheroids}
\label{sec:GEV_Field}

As noticed in Sect. \ref{sec:analysis}, star-forming spheroids represent a significant fraction (up to $25\%$) of the galaxies at the high-mass end ($M_\ast>10^{10.5}\,M_\odot$) in the field, decreasing with cosmic time up to $15\%$ in the lowest redshift bin ($0.25<z<0.5$). At the same time, the relative abundance of quiescent spheroids in the field and in the same mass range is found to increase (from $\sim30\%$ to $\sim50\%$). We couple these findings with the results presented in Sect.\,\ref{sec:MS} and shown in Fig.\,\ref{fig:main_sequence}, concerning the abundance of these galaxies in the green valley and at the high-mass end of the main sequence. In addition to these results, we recall that \citet{ArangoToro_25}, in their analysis of the evolutionary path of galaxies in the stellar mass versus SFR plane, found that the vast majority of the galaxies in the green valley were previously located at the high-mass end of the MS (between 70 and 90\% at $0.25<z<1$) and moves towards the red sequence because of a rapidly declining star-formation activity\footnote{We underline, however, that the study by \citet{ArangoToro_25} is only based on photometric observations. Hence, the reconstruction of non-parametric star formation histories might be affected by biases. Similar studies analysing spectroscopic data (e.g. \citealt{Mancini_19}) found higher fractions of rejuvenated galaxies in the green valley.}. All these findings clearly support a scenario where the bulges of star-forming galaxies grow during their evolution as part of the main sequence, together with the growth of stellar mass. Once galaxies reach the high-mass end of the MS, the {\rev crossing of the green valley and the transition to the red cloud are driven either by the simple ageing of the stellar population (e.g. \citealt{Bremer_18,Phillipps_19}) or by the quenching} (either mass- or morphology-driven; e.g. \citealt{Martig_09,Peng_10}).

This scenario is supported by several observational results. Firstly, the higher abundance of bulge-dominated galaxies at the high-mass end of the MS and in the green valley (see Fig.\,\ref{fig:main_sequence}, but also analogous studies by e.g. \citealt{Wuyts_11,Lang_14,Bremer_18,Huertas_24,Q1-SP031,Q1-SP040}). Secondly, the lower abundance of star-forming spheroids in denser environments, decreasing with cosmic time, even in the same mass range where they are more common in the field (see Fig.\,\ref{fig:peng_final}). In this case, most of these sources are the central galaxies of groups (as the vast majority of galaxies with $M_\ast>10^{10.5} \ M_\odot$ in dense environment; see Fig.\,\ref{fig:simulations}, but also analogous studies such as \citealt{McCracken_15,Popesso_19}) and -- therefore -- that the mass quenching efficiency is higher because of the larger halo masses (Sect. \ref{sec:simulations}), causing the earlier transition into the class of quiescent spheroids (see e.g \citealt{Peng_12}).

In terms of physical processes involved in our scenario for the formation of star-forming spheroids in the field, there is a rich literature of possible mechanisms able to increase the bulge size in star-forming galaxies. The main ones involve the accretion of small satellites via minor mergers (see e.g. \citealt{Bekki_11,Sachdeva_17,Tacchella_19}), through \textit{in-situ} star formation (see e.g. \citealt{Noguchi_99,Dekel_09b,Yu_22,Tan_24,Lyu_25}) {potentially resulting from gas accretion with misaligned angular momentum (see e.g. \citealt{Sales_12}) or clumps formation and migration through disc instabilities (see e.g. \citealt{Perez_13,Bournaud_16}}).  Similarly, the quenching of galaxies in the upper part of the MS can be explained by several mechanisms. A possible cause can reside in the presence of the bulge itself, as in the morphological quenching proposed by \citet{Martig_09}, where the bulge causes the stabilisation of the stellar disc and -- therefore -- the cessation of star formation. An alternative explanation can reside in the accretion of cold gas, expected to be inefficient in massive halos ($M_{\rm h}>10^{12} \ M_\odot$ in the redshift range covered by our observations) due to virial shock heating (see e.g. \citealt{Birnboim_03,Keres_05,Dekel_09a}). This phenomenon can easily explain the decrease of SFR for massive star-forming galaxies (e.g. \citealt{Daddi_22}). Finally, a role could be played by feedback mechanisms (e.g. by AGN), more common in massive galaxies above $10^{10.5}\,M_\odot$ (see e.g. \citealt{Bongiorno_16,Chen_20}) and with prominent bulges (e.g. \citealt{Ferrarese_00,Haring_04}). {A similar scenario would also be in agreement with what has been found by \citet{Correa_19} in their analysis of the EAGLE simulation \citep{Crain_15,Schaye_15}, where the time of maximum brightness of the SMBH in elliptical galaxies is found to tightly correlate with the time when the same objects cross the green valley, while such a correlation is negligible in discy galaxies.}

\subsection{Galaxy evolution in high-density environments and role of quiescent discs}
\label{sec:GEV_Dense}

In Sect. \ref{sec:analysis}, we noticed that the relative abundance of quiescent discs is almost negligible in the field, while it can reach fractions up to 20\% in over-dense regions. Moreover, these sources are always located at lower masses with respect to quiescent spheroids in dense environments, in a mass regime where we expect to be dominated by satellite galaxies (see Sect. \ref{sec:simulations} and Fig.\,\ref{fig:simulations}). At the same time, no significant trend with mass is visible in the field. Finally, their overall abundance increases with decreasing redshift (analogously to the strength of environmental effects).

We interpret all these results as evidence that quiescent discs are formed from star-forming discs through satellite quenching in dense environments. If some of them then stop their evolution, becoming the passive discs observed in the local Universe (e.g. \citealt{Masters_10}), others keep evolving transitioning into the class of quiescent spheroids. In more detail, in our scenario quiescent discs are prevented from forming stars because of external mechanisms in high-density environments (e.g. through tidal stripping or starvation, depending on the nature and state of the over-density; see e.g. \citealt{Meritt_83,Read_06,Feldmann_10}), lowering the amount of molecular gas available for star formation. Subsequently, other phenomena such as galaxy harassment and minor mergers causes the morphological transformation into bulge-dominated galaxies, eventually increasing their stellar mass. {Again, this scenario is in agreement with what has been found by \citet{Correa_19} in the EAGLE simulations, where quiescent discs are mainly found as satellites of groups and clusters and their morphology is transformed after the quenching of their star formation.}

\section{Summary}
\label{sec:summary}
In this paper, we characterised the galaxy quenching and the morphological transformation of disc-dominated galaxies into bulge-dominated ones as a function of mass and environment. For our analysis, we took advantage of the first data released by the Euclid Collaboration as part of its Q1 release, which covers about $60\,\mathrm{deg}^2$, reaching a limiting magnitude of $\IE=24.5$. After assembling a mass-complete sample of galaxies with $M_\ast>10^{9.5} \, M_\odot$ and characterising the density field through the $\Sigma_5$ estimator and the corresponding density contrast parameter $\log_{10}(1+\delta)$, we studied the evolution of the relative abundances of four families of galaxies as a function of stellar mass, local density, and redshift. These four classes of galaxies are based on the joint classification between star-forming and quiescent galaxies and disc- and bulge-dominated ones. {\rev The unprecedented statistics offered by the \Euclid telescope allowed us to assemble a statistical significant sample of galaxies belonging to the two intermediate populations of quiescent discs and star-forming spheroids and to study at the same time their distribution in mass and local density field. Our observational results can be summarised as follows.

\begin{itemize}
    \item Quiescent disc-dominated galaxies represent a negligible fraction of the sources in the field, where their relative abundance does not show any significant correlation with stellar mass. Conversely, these objects are significantly more common in denser environments, where their relative abundance is higher at lower masses and tends to increase with cosmic time.

    \item Star-forming bulge-dominated galaxies represent a significant fraction of the high-mass galaxies in the field, while their relative abundance at lower masses is negligible in all environments. Moreover, their relative fraction at the high-mass end of high-density environments tends to decrease with cosmic time. Finally, their relative abundance at the high-mass end of the field distribution tends to decrease with cosmic time, while an increasing fraction of quiescent bulge-dominated galaxies appears in the same region of the parameter space.
\end{itemize}

 These findings allow us to propose a simple evolutionary scenario, where the galaxy evolution in the field and in over-dense environments is significantly different. In more detail, our scenario is the following.}

\begin{itemize}

    \item In the field, the morphological transformation of galaxies happens mainly through secular processes taking place as part of the main sequence, with the formation of a dominant bulge once the stellar masses approach $10^{10.5} \, M_\odot$, where up to $25\%$ of the galaxies become star-forming bulge-dominated galaxies. {\rev The following migration to the red sequence happens thanks to internal processes. These can include the simple ageing of the stellar populations or the quenching, driven by possible mechanisms such as AGN feedback, morphological quenching, or virial shock heating driven by the large halo masses.}

    \item In higher-density environments, the evolution of structures follows the opposite trajectory, with the quenching of star formation taking place before the morphological transformation. In this scenario, the environmental quenching of star-forming discs (mostly satellite galaxies in large dark matter halos) produces quiescent discy galaxies, whose transformation into bulge-dominated quiescent galaxies eventually happens at a following stage through external mechanisms. {\rev These can include well-known processes such as dry mergers or galaxy harassment}.
    \end{itemize}

These results highlight the scientific potential of the \Euclid data in constraining the evolution of galaxies in different environments. These results are based on the first quick data release of the Euclid Consortium, covering a small fraction of the total area that will be observed in the wide survey and with a depth up to two magnitudes brighter than what will be reached in the deep survey at the end of the nominal length of the mission. On the one hand, the availability of a wider sky coverage will allow us to cover a larger dynamical range in the density contrast parameter, reaching levels where the environmental quenching is stronger even at higher redshifts. On the other hand, the availability of deeper data will allow us to extend our analysis to lower stellar masses and higher redshifts, to better constrain the photometric redshifts of our sources, and to better characterise their morphologies. In particular, the expected availability of bulge-disc decomposition for the galaxies included in the next data releases will allow us to better discriminate between disc- and bulge-dominated objects. Similarly, the forecasted improvement of the photometric redshifts in the next \Euclid data releases (see e.g. \citealt{EP-Enia}), will allow us to reduce the comoving size of our redshift slices and consequently the shot noise affecting the estimation of the density contrast. {A final improvement will be represented by the availability of slitless spectroscopy for most of the galaxies observed by \Euclid \citep{Q1-TP007,Q1-TP006}. These new data will - on the one hand - increase the spectroscopic coverage of our sample (still limited, see Sec. \ref{sec:specz}), and - on the other hand - allow us to identify remaining AGN. Although we expect our selection criteria, outlined in Section 2, to effectively minimise contamination of our sample by AGN host galaxies, some may still have been included, which potentially affected the physical properties derived by the Euclid pipeline (colours, SFR, and morphology).}

\begin{acknowledgements}
FaGe, AnEn, EmDa, LoGa, SaQu, GaDe, MaTa, ChDe, LuPo acknowledge support from the ELSA project. "ELSA: Euclid Legacy Science Advanced analysis tools" (Grant Agreement no. 101135203) is funded by the European Union. Views and opinions expressed are however those of the author(s) only and do not necessarily reflect those of the European Union or Innovate UK. Neither the European Union nor the granting authority can be held responsible for them. UK participation is funded through the UK HORIZON guarantee scheme under Innovate UK grant 10093177.
AnEn acknowledge support from the INAF MiniGrant 2023 "ADIEU: Anomaly Detections In EUclid". 
CaLo acknowledges support by FCT -Fundação para a Ciência e a Tecnologia through grants UIDB/04434/2020 DOI: 10.54499/UIDB/04434/2020, UIDP/04434/2020 DOI: 10.54499/UIDP/04434/2020.
\AckEC
\AckQone
\AckCosmoHub

Based on data from UNIONS, a scientific collaboration using three Hawaii-based telescopes: CFHT, Pan-STARRS, and Subaru \url{www.skysurvey.cc}\,. Based on data from the Dark Energy Camera (DECam) on the Blanco 4-m Telescope at CTIO in Chile \url{https://www.darkenergysurvey.org}\,.

\end{acknowledgements}

\bibliography{biblio}

\appendix
\onecolumn

\section{Properties of the density field}
\label{app:completeness}

\begin{figure*}
    \centering
    \includegraphics[width=\linewidth]{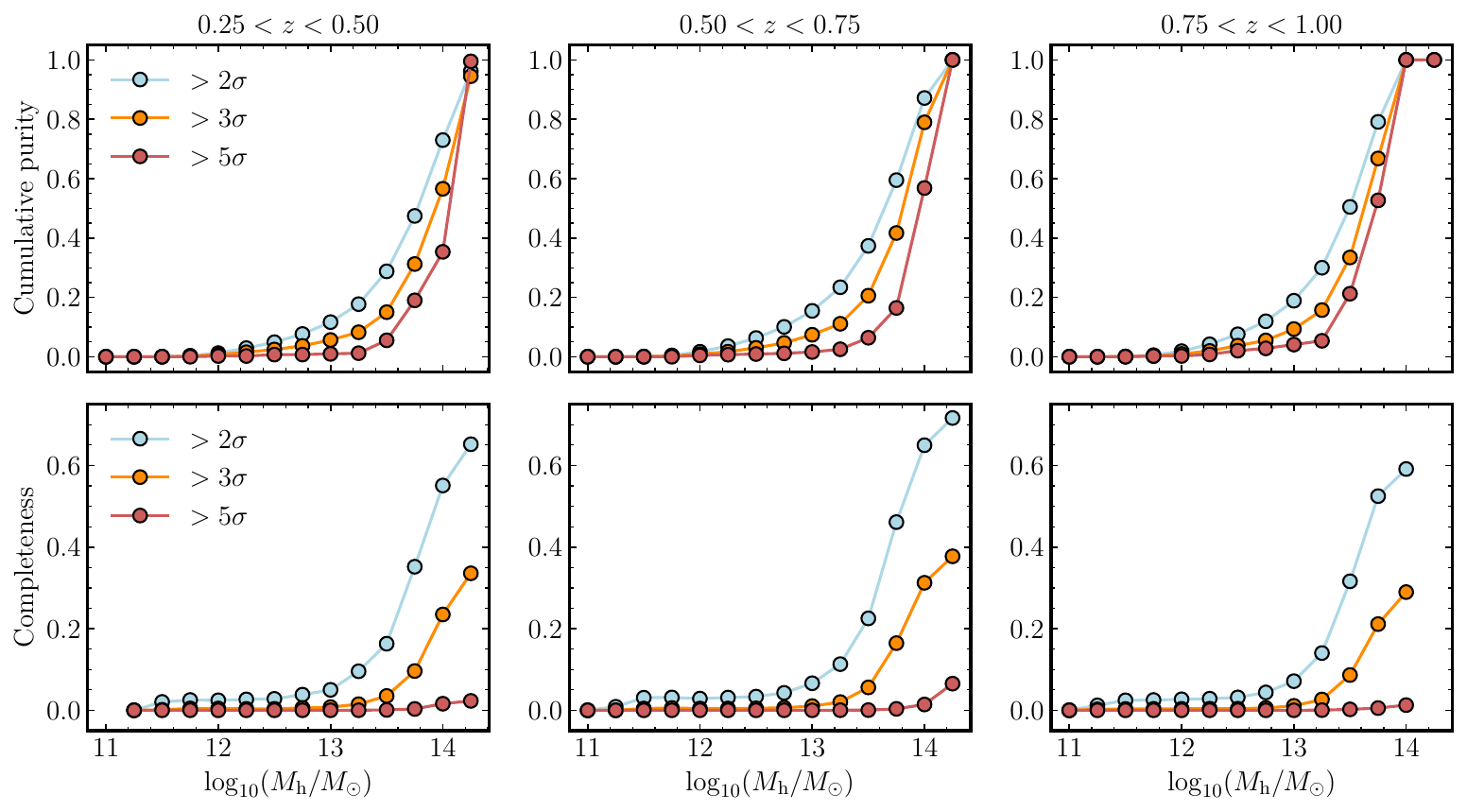}
    \caption{Cumulative purity and completeness of our estimation of the density field as a function of the halo mass and of the threshold employed to define over-dense regions, as assessed from the Euclid \texttt{Flagship-2} simulation.. A description of the two quantities can be found in Appendix \ref{app:completeness}.}
    \label{fig:completeness}
\end{figure*}

{\rev In order to estimate the reliability of our reconstruction of the density field in the EDFs, we apply the same pipeline described in Sect.\,\ref{sec:densty} to the simulated data from the \texttt{Flagship-2} simulation (see Sect.\,\ref{sec:simulations}). We adopt a quantitative approach by introducing two different metrics. The first is the cumulative purity (upper row of Fig.\,\ref{fig:completeness}): for a given value of the halo mass ($M_{\rm h}^*$), we define this quantity as the number of central galaxies with $M_{\rm h}<M_{\rm h}^*$ and with a density contrast $\log_{10}(1+\delta)$ above $n\sigma$ from the median divided by the total amount of central galaxies with $\log_{10}(1+\delta)$ above the same threshold. This quantity tells us what percentage of low-mass halos are included in our selection. From Fig.\,\ref{fig:completeness}, we can see how for all thresholds, more than 80\% of sources that -- according to our selection -- belong to over-densities actually have an halo mass compatible with a group ($M_{\rm h }>10^{13} \ M_\odot$).

The second metric that we employ is called completeness (lower row of Fig.\,\ref{fig:completeness}) and is defined, for a given value of the halo mass ($M_{\rm h}^*$), as the fraction of central galaxies with $M_{\rm h}=M_{\rm h}^*$ and with a density contrast above $n\sigma$ from the median, divided by the total amount of central galaxies with $M_{\rm h}=M_{\rm h}^*$. In other words, this quantity tells us what fraction of central galaxies with a given halo mass is recovered by our selection and -- therefore -- how complete is our sample of over-densities. From Fig.\,\ref{fig:completeness}, we can see how our selection is able to identify more than half of the massive structures ($M>10^{14} \ M_\odot$) when a threshold of $2\sigma$ is adopted, but how this fraction rapidly decreases when we choose a higher threshold or when we go to lower halo masses.

In summary, our selection of high-density regions is definitely pure (i.e. only structures with massive halos are retrieved by our selection), but not totally complete (several massive structure are missed by our selection). This result is easily explained by the employment of photometric redshift and, therefore, of thick redshift slices. This produces a shot noise due to the galaxies in the line-of-sight diluting the signal of real over-densities and biasing our selection towards the most massive structures.}

\end{document}